\documentclass[ArXiv,AXISWP,accept,moreauthors,pdftex,10pt,letterpaper]{axis} 

\firstpage{1} 
\pubvolume{xx}
\issuenum{1}
\articlenumber{5}
\pubyear{2023}
\copyrightyear{2023}

\usepackage{amsmath,amssymb}

\newcommand\arcsec{\hbox{$^{\prime\prime}$}}

\Title{Prospects for Time-Domain and Multi-Messenger Science with AXIS}

\Author{The AXIS Time-Domain and Multi-Messenger Science Working Group: Riccardo Arcodia$^{1,*}$, Franz E.~Bauer$^{2,3}$, S.~Bradley Cenko$^{4,5,\dagger}$, Kristen C.~Dage$^{6,*}$, Daryl Haggard$^{7,8,\dagger}$, Wynn C.~G.~Ho$^{9}$, Erin Kara$^{1}$, Michael Koss$^{10}$, Tingting Liu$^{11}$, Michela Negro$^{12,4,13}$, Pragati Pradhan$^{14,15}$, J. Quirola-V\'asquez$^{2,3,16}$, Mark T.~Reynolds$^{17,18}$, Richard E.~Rothschild$^{19}$, Navin Sridhar$^{20,21,22}$, Eleonora Troja$^{23}$, Yuhan Yao$^{24,25}$, Labani Mallick$^{26,27,22,\ddagger}$, Claudio Ricci$^{28}$}

\address{%
$^{1}$ \quad MIT Kavli Institute for Astrophysics and Space Research, Massachusetts Institute of Technology, Cambridge, MA 02139, USA \\
$^{2}$ \quad Instituto de Astrof\'{\i}sica, Facultad de F\'{i}sica, Pontificia Universidad Cat\'{o}lica de Chile, 306, Santiago 22, Chile \\
$^{3}$ \quad Millennium Institute of Astrophysics (MAS), Nuncio Monse\~{n}or S\'{o}tero Sanz 100, Providencia, Santiago, Chile \\
$^{4}$ \quad Astrophysics Science Division, NASA Goddard Space Flight Center, Greenbelt, MD 20771, USA \\
$^{5}$ \quad Joint Space-Science Institute, University of Maryland, College Park, MD 20742, USA \\
$^{6}$ \quad Wayne State University, Department of Physics \& Astronomy, 666 W Hancock St, Detroit, MI 48201, USA \\
$^{7}$ \quad Trottier Space Institute at McGill, 3550 University Street, Montreal, QC H3A 2T8, Canada \\
$^{8}$ \quad Department of Physics, McGill University, 3600 University Street, Montreal, QC H3A 2A7, Canada \\
$^{9}$ \quad Department of Physics and Astronomy, Haverford College, 370 Lancaster Avenue, Haverford, PA 19041, USA \\
$^{10}$ \quad Eureka Scientific, 2452 Delmer Street Suite 100, Oakland, CA 94602-3017, USA \\
$^{11}$ \quad Department of Physics and Astronomy, West Virginia University, P.O. Box 6315, Morgantown, WV, 26506, USA \\
$^{12}$ \quad University of Maryland, Baltimore County, Baltimore, MD 21250, USA \\
$^{13}$ \quad Center for Research and Exploration in Space Science and Technology, NASA/GSFC, Greenbelt, MD 20771, USA \\
$^{14}$ \quad Massachusetts Institute of Technology, 77 Massachusetts Avenue, Cambridge, MA 02139, USA \\
$^{15}$ \quad Embry Riddle Aeronautical University, Department of Physics \& Astronomy, 3700 Willow Creek Road, Prescott, AZ 86301, USA \\
$^{16}$ \quad Observatorio Astron\'omico de Quito, Escuela Polit\'ecnica Nacional, 170136, Quito, Ecuador \\
$^{17}$ \quad Department of Astronomy, University of Michigan, 1085 South University Avenue, Ann Arbor, MI 48109, USA \\
$^{18}$ \quad Department of Astronomy, Ohio State University, 140 W 18th Avenue, Columbus, OH 43210, USA \\
$^{19}$ \quad Department of Astronomy and Astrophysics, University of California at San Diego, La Jolla, CA 92093-0424, USA \\
$^{20}$ \quad Department of Astronomy, Columbia University, New York, NY 10027, USA \\
$^{21}$ \quad Theoretical High Energy Astrophysics (THEA) Group, Columbia University, New York, NY 10027, USA \\
$^{22}$ \quad Cahill Center for Astrophysics, California Institute of Technology, MC 249-17, 1200 E California Boulevard, Pasadena, CA 91125, USA \\
$^{23}$ \quad Department of Physics, University of Rome -- Tor Vergata, via della Ricerca Scientifica 1, I-00100, Rome, Italy \\
$^{24}$ \quad Miller Institute for Basic Research in Science, 468 Donner Lab, Berkeley, CA 94720, USA \\
$^{25}$ \quad Department of Astronomy, University of California, Berkeley, CA 94720, USA \\ 
$^{26}$ \quad Department of Physics \& Astronomy, University of Manitoba, Winnipeg, Manitoba R3T 2N2, Canada\\
$^{27}$ \quad Canadian Institute for Theoretical Astrophysics, University of Toronto, 60 St George Street, Toronto, Ontario M5S 3H8, Canada\\
$^{28}$ \quad Instituto de Estudios Astrof\'{\i}sicos, Facultad de Ingenier\'{\i}a y Ciencias, Universidad Diego Portales, Avenida Ejercito Libertador 441, Santiago, Chile \\
$^{\dagger}$ \quad AXIS TDAMM Science Working Group Co-Lead \\
$^*$ \quad NASA Einstein Fellow \\
$^{\ddagger}$ \quad CITA National Fellow \\

}

\abstract{The Advanced X-ray Imaging Satellite (AXIS) promises revolutionary science in the X-ray and multi-messenger time domain. AXIS will leverage excellent spatial resolution ($<1.5$ arcsec), sensitivity ($80\times$ that of Swift), and a large collecting area ($5-10\times$ that of Chandra) across a 24-arcmin diameter field of view to discover and characterize a wide range of X-ray transients from supernova-shock breakouts to tidal disruption events to highly variable supermassive black holes. The observatory's ability to localize and monitor faint X-ray sources opens up new opportunities to hunt for counterparts to distant binary neutron star mergers, fast radio bursts, and exotic phenomena like fast X-ray transients. AXIS will offer a response time of $<2$ hours to community alerts, enabling studies of gravitational wave sources, high-energy neutrino emitters, X-ray binaries, magnetars, and other targets of opportunity. This white paper highlights some of the discovery science that will be driven by AXIS in this burgeoning field of time domain and multi-messenger astrophysics.
\emph{This White Paper is part of a series commissioned for the AXIS Probe Concept Mission; additional AXIS White Papers can be found at the  \href{http://axis.astro.umd.edu/}{AXIS website} with a mission overview \href{https://arxiv.org/abs/2311.00780}{here}}.
}

\begin{document}

\tableofcontents
\listoffigures

\section{Introduction}
Time-domain and multi-messenger astronomy (TDAMM) are highlighted as one of the three priority science areas in the coming decade by the 2020 Decadal Survey. In large part, this is the result of the dramatically improved sensitivity of foundational facilities across the electromagnetic spectrum (e.g., Vera Rubin Observatory \cite{2019ApJ...873..111I}, Nancy Grace Roman Space Telescope \cite{2015arXiv150303757S}, Square Kilometer Array \cite{2009IEEEP..97.1482D}) and beyond (e.g., LIGO-Virgo-KAGRA \cite[LVK; ][]{2020LRR....23....3A} and IceCube \cite{IceCubeGen2}). To fully realize the potential of TDAMM in this new era, the community needs electromagnetic observatories that can match these tremendous sensitivity gains. With an order of magnitude increase in effective area over Chandra and excellent angular resolution across a wide field of view, the Advanced X-ray Imaging Satellite (AXIS; \cite{AXISOverview}) can uniquely fill this void at X-ray wavelengths.

Building on the legacy of facilities such as Swift, Chandra, and XMM-Newton, AXIS TDAMM studies will address some of the most pressing topics in astrophysics. This includes investigations conducted by the science team and a robust target-of-opportunity (ToO) program for Guest Investigator (GI) programs. In this white paper, we briefly describe some of the diverse areas of TDAMM that will be transformed by AXIS in the next decade.

\section{Fast X-ray Transients}\label{sec:FXT}
Fast X-ray transients (FXTs) are single bursts (i.e., not related to known persistent X-ray sources) of X-ray photons that last from minutes to hours \citep{Heise_FXT_2010}. Historically, most detected FXT candidates have occurred along the Galactic plane and are Galactic in origin \citep[e.g.,][]{Pye_FXT_1983,Arefiev_FXT_2003,Heise_FXT_2010}. However, a subset lies well outside the Galactic plane and thus is potentially extragalactic \citep[e.g.,][]{Ambruster_FXT_1986,Hurley_FXT_1999}. 
Extragalactic FXTs are particularly interesting because of their potential energetics, rarity, and association with exotic phenomena. 

Several dozen extragalactic FXTs have been identified, both serendipitously and through careful searches
\citep[e.g.,][]{Soderberg_FXT_2008,Jonker_FXT_2013,Glennie_FXT_2015,Bauer_FXT_2017,Lin_FXT_2018,Lin_FXT_2019,Xue_FXT_2019,Alp_FXT_2020,Novara_FXT_2020,Lin_FXT_2020,Lin_FXT_2021,Lin_FXT_2022,Eappachen_FXT_2022,Eappachen_FXT_2023,Quirola_FXT_2022,Quirola_FXT_2023}. These observations suggest that extragalactic FXTs may be associated with several novel classes of astronomical objects \citep[e.g.,][]{Maguire_FXT_2020,Saxton_FXT_2021,Bayless_FXT_2022}. The most well-known case is FXT XRT~080109/SN~2008D, which was serendipitously detected by Swift-XRT and subsequently associated with a multi-wavelength supernova counterpart \citep[][]{Soderberg_FXT_2008,Mazzali_FXT_2008,Modjaz_FXT_2009}. However, for the vast majority of cases, transients themselves have only been identified long after the outburst through archival data mining \citep[e.g.,][]{Alp_FXT_2020,DeLuca_FXT_2021,Quirola_FXT_2022,Quirola_FXT_2023}, leaving contemporaneous follow-up observations largely unexplored. 
This has left the task of identifying the physical origin(s) of FXTs quite challenging.

A variety of astronomical objects and physical mechanisms have been proposed for the
origin of extragalactic FXTs.
These include strong sources of high- and low-frequency gravitational waves (GWs) in the form of merging binary neutron stars \citep[BNS; potential LVK GW sources; e.g.,][]{Kramer_FXT_2006,Zhang_FXT_2013} or white dwarf (WD) disruptions by intermediate-mass black holes \citep[potential future LISA GW observatory sources; e.g.,][]{Sesana_FXT_2008,Maguire_FXT_2020}, 
as well as core-collapse supernova (CC-SNe) shock breakouts \citep[SBOs; e.g.,][]{Waxman_FXT_2017,Alp_FXT_2020}. Thus, exploring their origin has significant potential implications across several fields (see Sections~\ref{SM-BNS}, \ref{SNe}, and \ref{TDE} for more details). 

Efforts have been made to identify, classify, and characterize FXTs inside Chandra \citep[e.g.,][]{Lin_FXT_2022,Quirola_FXT_2022,Quirola_FXT_2023}, XMM-Newton \citep[e.g.,][]{Alp_FXT_2020,DeLuca_FXT_2021}, and Swift-XRT \citep{Evans_FXT_2023} archives. 
Two decades of archival Chandra data have revealed 22 FXTs 
\cite{Quirola_FXT_2022,Quirola_FXT_2023}: 
5 events were robustly associated with galaxies at ${\lesssim}$100~Mpc (the \emph{local sample}) and 17 events appear to lie much further away at ${\gtrsim}$100~Mpc (the \emph{distant sample}). The \emph{local sample} has a peak luminosity of $L_{\rm X,peak}^{\rm Local}{\lesssim}10^{40}$~erg~s$^{-1}$, an event rate of $\mathcal{R}_{\rm Local}{=}$53.7$_{-15.1}^{+22.6}$~deg$^{-2}$~yr$^{-1}$ \citep{Quirola_FXT_2022,Quirola_FXT_2023} and projected physical offsets between ${\approx}$0.7 and 9.4~kpc (with four being co-spatial with apparent star-forming regions or young star clusters). These properties indicate a possible association of the \emph{local sample} of FXTs with ultra-luminous X-ray sources or X-ray binaries (e.g., Sections~\ref{sec:magnetars} and \ref{sec:XRBs}).
On the other hand, the \emph{distant sample} properties show a link with energetic progenitors ($L_{\rm X,peak}^{\rm Distant}{\gtrsim}10^{40}$~erg~s$^{-1}$) at higher distances ($z{\approx}$0.3--2.2) with an event rate of $\mathcal{R}_{\rm Distant}{=}$36.9$_{-8.3}^{+9.7}$~deg$^{-2}$~yr$^{-1}$. Their high luminosities suggest a variety of origins such as BNS mergers ($L_{\rm X,peak}^{\rm BNS}{\approx}10^{44}-10^{46}$~erg~s$^{-1}$;  \citep{Metzger_FXT_2014,Sun_FXT_2017,Sun_FXT_2019,Metzger_FXT_2018b}), tidal disruption events involving IMBH ($L_{\rm X,peak}^{\rm TDE}{\lesssim}10^{48}$~erg~s$^{-1}$; \citep{Bloom_FXT_2011,MacLeod_FXT_2014,Maguire_FXT_2020}), and CC-SNe SBOs ($L_{\rm X,peak}^{\rm SBO}{\approx}$10$^{42}$--10$^{45}$~erg~s$^{-1}$;  \citep{Soderberg_FXT_2008,Modjaz_FXT_2009,Waxman_FXT_2017,Alp_FXT_2020,Novara_FXT_2020}).

\begin{figure}[t]
    \centering
    \includegraphics[scale=0.70]{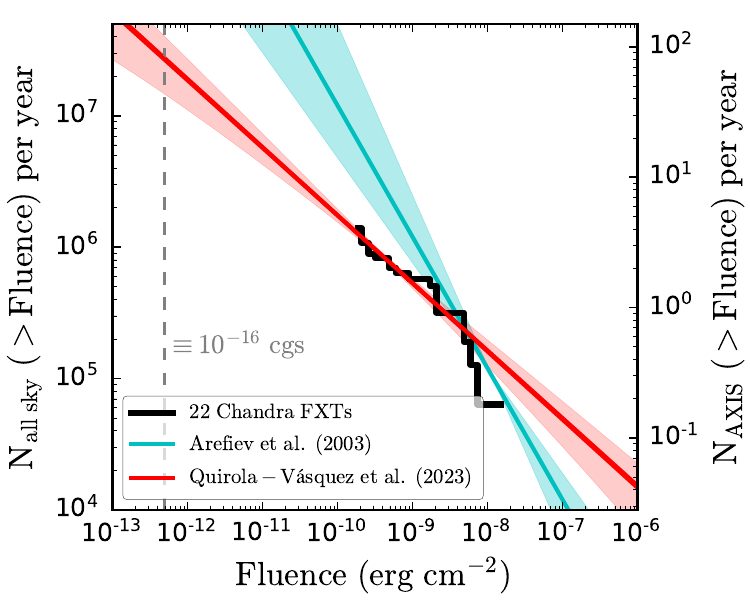}
    \caption[Expected log$\mathcal{N}$--log$S$ distributions of FXTs]
    {Expected log$\mathcal{N}$--log$S$ distributions of FXTs, all-sky (left vertical axis) and with AXIS (right vertical axis), based on combined constraints from several past missions \citep{Arefiev_FXT_2003} and 22 years of Chandra data \citep{Quirola_FXT_2023}, which constrains the break at a fluence of ${\approx}5\times10^{-9}$~erg\,cm$^{-2}$. AXIS will probe the FXT number density 2.5\,dex deeper than existing data, down to an approximate single-orbit AXIS flux sensitivity (dashed vertical line), greatly increasing our understanding of the FXT population.
    \label{fig:FXT_rate}
    }
\end{figure}

We expect FXT progenitors to form around some underlying stellar population -- a host galaxy -- which provides a route to measuring the event's distance and energetics. The demographics and offsets of the hosts themselves can further inform progenitor models. SBOs should arise from star-forming galaxies across a wide range of stellar masses, from blue and compact dwarf galaxies to large spiral galaxies \citep{Taggart_FXT_2021}. 
Similarly, long GRBs are preferentially associated with irregular star-forming galaxies, along with a few spirals with active star-formation \citep{Fruchter_FXT_2006}. Their host galaxies are relatively metal-poor compared to the field population \citep{Fynbo_FXT_2003,Prochaska_FXT_2004,Fruchter_FXT_2006}, favoring a collapsar progenitor model \citep{Woosley_FXT_2006}. On the other hand, IMBH TDEs involving WDs might occur in a more diverse range of environments such as irregular dwarf galaxies, globular clusters, and hyper-compact stellar clusters \citep[e.g.,][]{Merritt_FXT_2009,Jonker_FXT_2012,Reines_FXT_2013}, resulting in substantial offsets from the center of their host galaxy. Recently, efforts have been made to identify FXT host galaxies (e.g., XRT~000519, XRT~030511, and XRT~210423) using optical and near-infrared 6-, 8- and 10-meter telescopes \citep{Eappachen_FXT_2022,Eappachen_FXT_2023,Lin_FXT_2022}, but the hosts 
are optically faint ($r{\gtrsim}23$~AB~mag), and hence accurate X-ray positions are essential to make firm associations.

The angular resolution accuracy afforded by Chandra, XMM-Newton, and Swift-XRT has permitted to pinpoint hosts and even allow for measurements of apparent offsets to their potential host galaxies. The projected physical offset between the FXT position and the host galaxy center gives some clues about their nature. For example, short GRBs have a physical offset about ${\sim}$4--5 times greater than the median offset for long GRBs (${\approx}1$~kpc) \citep{Bloom_FXT_2002} and super-luminous supernovae (${\approx}15$~kpc) \citep[SL-SNe]{Schulze_FXT_2021}, and about ${\sim}$1.5 times larger than the median offsets of CC- and Type Ia SNe (${\approx}3-5$~kpc) \citep{Prieto_FXT_2008} and FRBs (${\approx}3$~kpc) \citep{Heintz_FXT_2020}. 
Thus, the observed offset distribution of short GRBs agrees with the population synthesis models for compact object mergers, especially for large offsets. The high accuracy angular resolution of AXIS (${\approx}1$~arcsec) will permit associating the FXT position with its host galaxy and measuring its angular offset.

\begin{figure}[t]
    \centering
    \includegraphics[width=1.0\columnwidth]{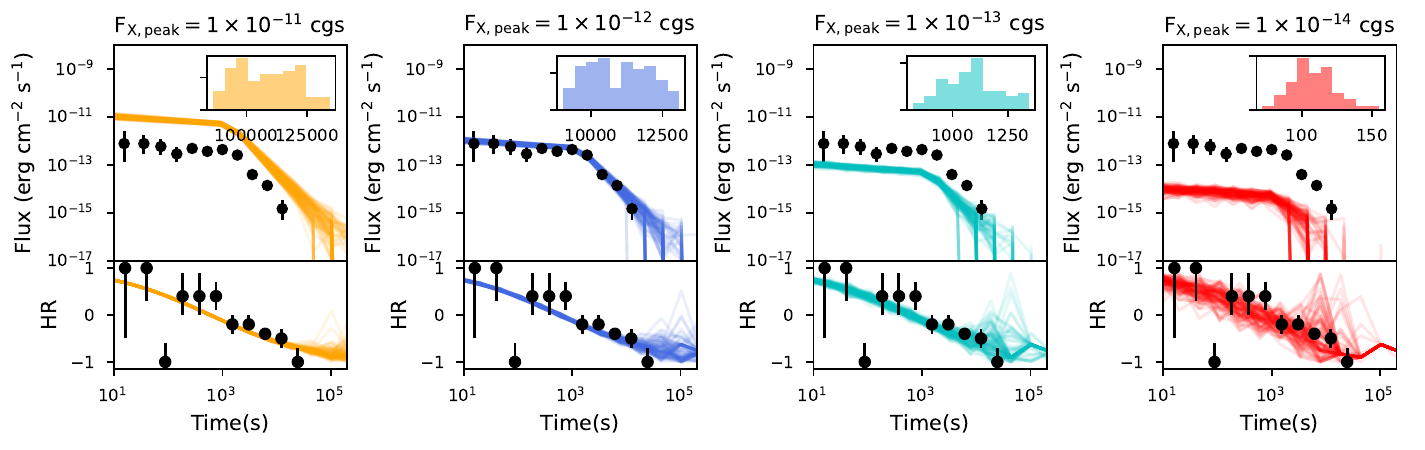}
    \caption
    [Simulated AXIS light curves and hardness ratios of FXTs]
    {Simulated AXIS light curves (upper panels) and hardness ratios (lower panels) of FXTs with 5\,ks 0.5--8\,keV peak fluxes of 1$\times$10$^{-11}$, 1$\times$10$^{-12}$, 1$\times$10$^{-13}$, and 1$\times$10$^{-14}$ erg\,s$^{-1}$\,cm$^{-2}$, assuming the light-curve properties of CDF-S XT2 \citep[Chandra light curve shown as black points;][]{Xue2019}. We adopt a hardness ratio given by H-S/H+S, where S=0.5--2 keV and H=2--8 keV. The colored light curves denote 100 AXIS realizations to demonstrate the error distribution. The inset histogram shows the number of counts for the simulated AXIS light curves; the real AXIS light curve will also incur some gaps due to the low-Earth orbit. AXIS will be capable of probing subtle spectral variations and temporal structure for many dozens of bright FXTs, and will broadly characterize the flux and spectral evolution for hundreds of faint FXTs.
    \label{fig:FXT_LC_spec}}
\end{figure}

Based on extrapolation of the current FXT statistics (Figure~\ref{fig:FXT_rate}), we anticipate that AXIS will serendipitously discover $\approx$50 FXTs yr$^{-1}$, 
representing a $\gtrsim$30-fold increase over current samples, as well as provide crucial windows for follow-up of bright FXT triggers from all-sky monitors \citep{Quirola_FXT_2023}.
The novel grasp (FOV $\times$ sensitivity) and spatial resolution of AXIS will provide better photon statistics with which to characterize each FXT, spectrally and temporally, precise locations to pin down host galaxy identifications and FXT locations/offsets, and rapid notifications to enable broad multi-wavelength follow-up campaigns. Figure~\ref{fig:FXT_LC_spec} shows representative light curves and hardness ratios at bright, moderate, and faint flux thresholds; for the brightest 5--10 FXTs detectable per year, AXIS will provide time-resolved spectral properties 
and accumulate powerful statistics for each proposed progenitor class.

\section{Gravitational-Wave Counterparts}

\subsection{Stellar-Mass Compact Binary Mergers}\label{SM-BNS}
\begin{figure}[t]
    \centering
    \includegraphics[width=0.7\columnwidth]{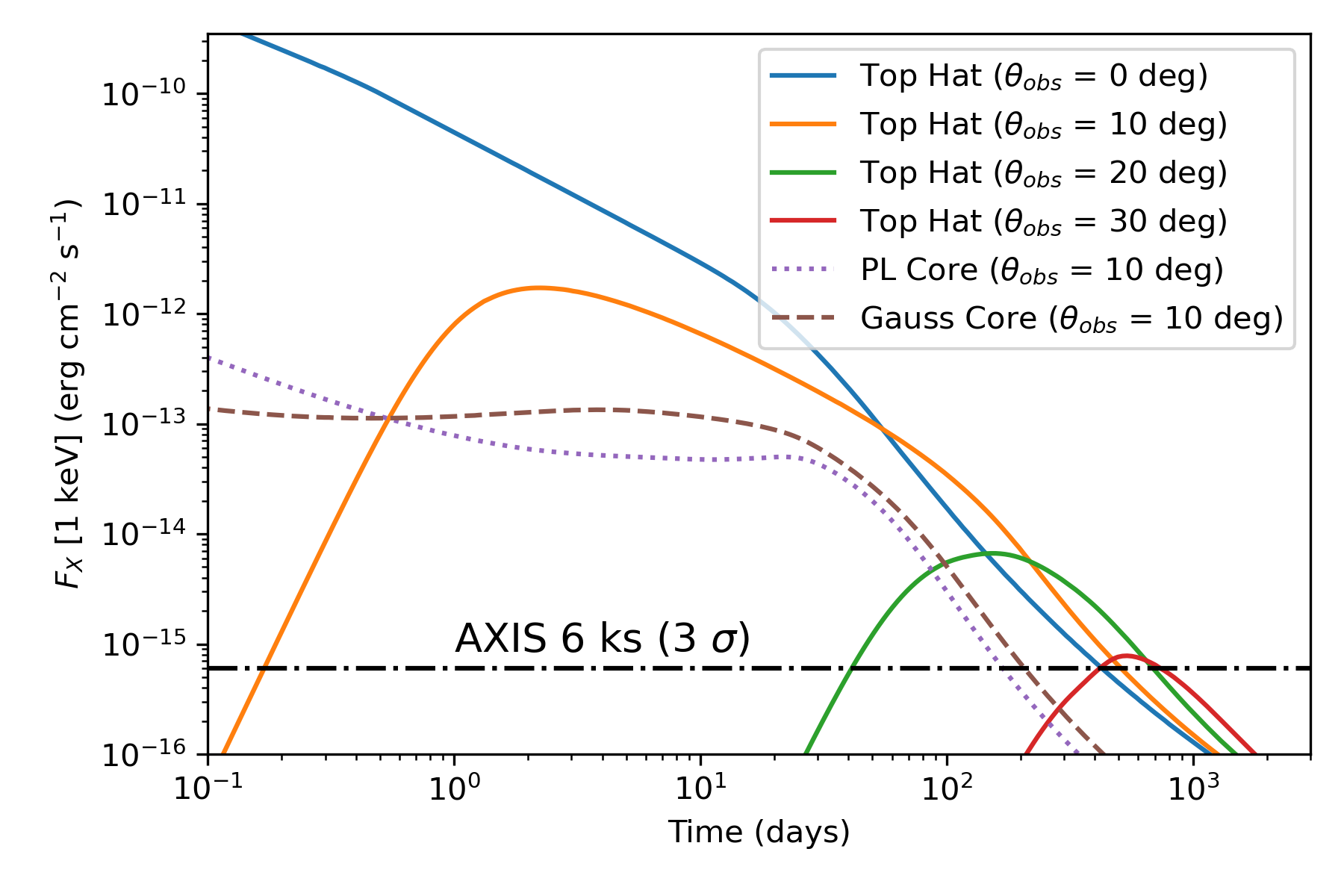}
    \caption
    [Jet model comparison for a variety of BNS merger observing angles]
    {Top-hat jet models (15 deg opening angle, $\Gamma = 2.2$) for a variety of BNS merger observing angles (0, 10, 20 and 30 degrees), compared to two alternative structured jets with a Powerlaw (PL) or Gaussian core, each at a 10 degree observing angle. The black dot-dashed line indicates the AXIS sensitivity reached in 6 ks. All models assume a luminosity distance $d_L = 330$ Mpc, isotropic-equivalent energy $E_0 = 10^{53}$ erg, and circumburst density $10^{-3}$ cm$^{-3}$. The strength of the X-ray signal at peak depends strongly on observing angle, while the shape of the lightcurve reveals the relativistic structure of the outflow. Models created using the open source tool {\tt afterglowpy} \cite{Ryan+2020}.
    \label{fig:BNS_LCs}}
\end{figure}

Colliding neutron stars emit bursts of electromagnetic and gravitational radiation, each providing unique insights into the physics of the merger and its ability to drive relativistic outflows. GW detections yield distances, progenitor masses, and spins, while relativistic jets and afterglows detected via X-ray and other multi-wavelength observations reveal the 
off-axis angle and the conditions of the surrounding interstellar medium (e.g., \cite{Troja+2017,Ryan+2020,Margutti+2021}). Potential precursors to gravitational wave events could also shed light on the pre-destruction properties of the neutron star, such as its magnetic field strength \citep{Sridhar+21}. Understanding the inclination is key for using binary neutron stars as cosmological standard candles, as the greatest uncertainty in the gravitational wave distance is its degeneracy with inclination (e.g., \cite{Nissanke+2013,Abbott+2017b}). Moreover, X-ray observations powerfully discern the structure of the relativistic jet and the merger remnant (Figure \ref{fig:BNS_LCs}), and offer a closer look at kilonova afterglows, outflows that trace the energetics of the central explosion and constrain the synthesis of heavy elements. AXIS will obtain light curves and spectra for more than 50 BNS mergers, 
creating the first X-ray population study of multi-messenger sources.

GW experiments have rapidly made new discoveries \cite{Abbott+2020a}, including the first definitive detection of a neutron star merger, GW170817 (e.g., \cite{Abbott+2017a}).
These explosions reveal our cosmic chemistry by pinpointing the synthesis and abundance of the heaviest elements of the periodic table: rapid neutron capture ($r$-process) elements. They also offer some of our best constraints on the conditions at the core of a neutron star (e.g., \cite{Abbott+2018}). Yet to date we have only one example of a GW+EM detection of one of these mergers, GW170817. In the LVK A+ era (observing run 5 [O5] and beyond), the detection range for BNS mergers will be $\sim$325 Mpc \cite{2020LRR....23....3A},
yielding counterparts that can only be detected by the most sensitive X-ray and radio instruments. (LIGO A+ is fully funded and already being implemented and LIGO-India is proceeding well and expected to join the network in 2030.) Detailed studies of the GW170817 structured jet and afterglow (e.g., \cite{Morsony+2023})
suggest that an X-ray instrument with the sensitivity of AXIS could detect $\geq 10$ GW170817-like afterglows per year ($>20$\% of events even beyond 1\,Gpc). X-rays would thus offer a census of jets and afterglows from GW sources.  

The exquisite spatial resolution offered by AXIS is also key to distinguishing BNS mergers from variable nuclear X-ray sources in the host galaxy. At $\sim$40 Mpc GW170817 was separated by only 10 arcsec from the low-luminosity AGN in NGC 4993. AXIS’s PSF will be required to resolve similar systems out to $\sim$300--400 Mpc, where most future gravitational wave discoveries will be made in the LVK A+ era. GW170817-based models indicate that only 16 additional binary neutron star mergers with high-quality jet angle determinations would constrain $H_0$ to better than 2\%, vs. the 50 to 100 sources that would be needed without afterglow measurements (e.g., \cite{Hotokezaka+2019}). The high sensitivity and spatial resolution offered by AXIS, particularly at distances beyond 200 Mpc, are thus essential to 
realizing the promise of multi-messenger constraints in cosmology.  

During its primary mission, AXIS observations of more than 50 BNS mergers will be triggered by a GW detection from LIGO-Virgo-KAGRA and a rapidly localized electromagnetic counterpart from one of several next-generation surveys, e.g., Vera Rubin Observatory’s LSST, BlackGEM, DECam, Roman Space Telescope, and many others. These exceptional new facilities are optimized for efficient tiling of the gravitational wave localization region (typically $10-100$ deg$^2$ or larger) and will offer rapid public alerts. AXIS 
observations will 
support detailed modeling of the relativistic jet, yielding   constraints on the jet geometry and propagation, the nature of the merger remnant (either a BH or a massive NS), 
and the associated rates for sGRBs and NS mergers (including both NS-NS and NS-BH).

\subsection{Solitary and non-merging neutron stars}
Another important type of GW source for which AXIS observations could prove crucial is isolated neutron stars and neutron stars in non-merging systems. These neutron stars can produce detectable persistent GWs if they have, e.g., large mountains, magnetic fields, or fluid oscillations (see, e.g., \cite{Riles2023,Wette2023}, for reviews).  Many of the best candidates for such GW emission are young neutron stars that are bright in X-rays but not necessarily bright in radio, and many of these are surrounded by diffuse X-ray emission from supernova remnants and pulsar wind nebulae and/or in crowded fields.  AXIS could identify potential new candidates as well as provide observations that enable the most sensitive GW searches.  Finally, the $\sim$arcsecond spatial resolution of AXIS would be vital to confirm sources detected first by GWs and thus enable follow-up observations at other electromagnetic wavelengths.

\subsection{(Super)massive Black Hole Binaries and Mergers}
Supermassive black hole binaries (SMBHBs) are thought to be the natural outcomes of the standard framework of hierarchical structure formation (e.g., \citep{Begelman1980}): the merger of galaxies brings their central SMBHs to the nucleus of the newly merged system, initially forming a \emph{dual} SMBH (or dual AGN if both BHs are active) and later becoming a gravitationally bound \emph{binary} at a separation where the mass enclosed within the orbit is less than the BH mass. Modern numerical simulations show that a gaseous circumbinary disk can form around a close-separation SMBHB (e.g., \citep{MacFadyen2008,Shi2012,Noble2012,D'Orazio2013}) and efficiently deliver gas to the BHs through a pair of narrow streams that feed their individual accretion disks (so-called ``minidisks''; e.g., \citep{Farris2014,Tang2017,Bowen2018,Paschalidis2021}). This process could sufficiently power an SMBHB to radiate as a binary AGN observable across the electromagnetic spectrum. 

The science of SMBHBs is multifaceted. It holds the key to our understanding of the role of mergers in SMBH growth and the evolution of SMBHs in the context of their host galaxies. As binary AGN, SMBHBs are laboratories for understanding gas dynamics and accretion physics in time-evolving spacetimes. The distinctive binary disk structure predicted in numerical simulations may be tested through the observations of peculiar AGN spectral features: for instance, the source may show X-ray spectral hardening as the result of streams striking the minidisks \citep{Roedig2014,Farris2015}, or a double Fe K$\alpha$ line originating from the minidisks (e.g., \citep{Sesana2012}). As the binary's orbital motion can often be imprinted on the AGN flux as a periodic variation, either as the result of binary-modulated accretion (e.g., \citep{MacFadyen2008,Shi2012,Noble2012,D'Orazio2013,Farris2014,Gold2014}), relativistic Doppler boosting \citep{D'Orazio2015}, gravitational lensing \citep{D'Orazio2018selflensing,Davelaar2022}, or other mechanisms (e.g., \citep{Bowen2017,Tang2018,Ivanov1998,Lehto1996}), binary AGN are also interesting sources to study for time-domain astronomy. The predicted EM and time-domain signatures of these sources, and future prospects with AXIS, are discussed in detail in the associated AXIS white paper ``Tracking SMBH Mergers from kpc to Sub-pc Scales with AXIS.''


Observations of SMBHBs are also highly synergistic with low-frequency GW detectors, including pulsar timing arrays (PTAs) and the Laser Interferometer Space Antenna (LISA), and will together enable a new area of multi-messenger astrophysics. A global consortium of PTA experiments, including the US-led North American Nanohertz Observatory for Gravitational Waves (NANOGrav), has recently revealed evidence for a nanohertz frequency GW background \citep{NG15yrGWB,Reardon2023,Antoniadis2023,Xu2023} whose amplitude and spectral shape are consistent with a population of SMBHBs \citep{NG15yrAstro,Antoniadis2023Astro}. If this background indeed arises from SMBHBs, the loudest among them could be detected as single sources as early as $\sim$ 2030 \citep{Kelley2018SMBHB,Rosado2015}, thanks to powerful new workhorse radio facilities such as the Deep Synoptic Array-2000 (DSA-2000; \cite{Hallinan2019}) and the SKA \cite{Lazio2013} (see, e.g., \cite{Liu2023}).

Once the SMBHB is detected in GWs and localized within a sky area, telescopes can then be deployed to identify its EM counterpart, for instance, by searching for AGN periodicity indicative of an SMBHB, X-ray spectral hardening or excess, or an oscillating double broad Fe line. Conversely, the sky location and binary parameters of an EM-detected SMBHB can be used as priors in the search in PTA data for a ``GW counterpart.'' These types of joint multi-messenger observations yield higher signal-to-noise ratios and tighter parameter constraints \citep{Liu2021}, and therefore they can distinguish marginal sources that would otherwise be missed in unguided, GW-only searches. The more precise parameter measurements can further break model degeneracies (such as the mass-ratio dependence of binary periodicities) and enable stringent tests of the theory of binary accretion.

Starting in the mid-2030s, the space-based LISA mission \citep{Amaro-Seoane2017} will operate in the mHz frequency range and detect, among many other types of sources, the mergers of massive black holes (MBHs) in the $\sim 10^{5}-10^{7} M_{\odot}$ range. During its nominal 4-year science operations, LISA is expected to detect a few dozen to a few hundred MBHBs (e.g., \citep{Mangiagli2022,Barausse2023}) and can localize these sources and constrain their parameters $\sim$ hours to weeks before the merger (e.g., \citep{Mangiagli2020}), providing advance warning for EM follow-up. 

A sensitive X-ray telescope such as AXIS will be particularly powerful for probing MBHBs at this stage, since X-rays trace gas in the immediate vicinity of the binary (i.e., minidisks), whereas the source may be indistinguishable from a single AGN in the optical band (which is dominated by gas further out). In these systems, the binary separations are small enough for general relativistic effects to be significant (e.g., \citep{Bowen2017,Bowen2019}); observations of these LISA systems would therefore offer opportunities to probe binary accretion in a (relativistic) regime that is not accessible with PTA counterparts. An agile telescope like AXIS will be able to respond quickly to a LISA trigger, offering the best chance to catch the merger in the act. The sky localization area of a LISA source depends on its parameters (e.g., mass) and redshift but is likely large ($\sim$10 -- 10$^{2}$ deg$^{2}$), with a strong trade-off between localization uncertainty and time until merger (e.g., \citep{Mangiagli2020}). Although the FOVs of X-ray telescopes are small compared to optical (e.g., Rubin) and radio (e.g., SKA) facilities, follow-up in X-rays will still be crucial because the optical band is more susceptible to obscuration, and radio emission could be collimated. Furthermore, targeting X-ray-selected AGN reduces the number of potential hosts within the LISA localization area by orders of magnitude \cite{Lops2023}, increasing the likelihood of identifying the counterpart before the merger.

At the time of the merger, the LISA localization error box may be $\sim 0.1$ deg$^{2}$ \citep{Piro2023}, which would fit comfortably inside the AXIS FOV; this provides opportunities to observe the post-merger prompt or delayed emission and to witness the birth of a new (single) AGN. \citet{Yuan2021} predict that the jet launched after the merger of an MBHB pushes through the disk wind material originating from the (former) circumbinary disk and minidisks, and the resulting broadband emission is observable $\sim$ days--months after the merger. Additionally, gravitational recoil at merger imparts a kick velocity of $\sim$ a few hundred km s$^{-1}$ on the disk, which may produce a transient flare over timescales of $\sim$ years \citep{Rossi2010}. Many of these mergers could be associated with lower mass systems at higher redshifts, and detecting their post-merger emission would therefore require deep imaging. A 10$^{5}M_{\odot}$ system at $z=2$ accreting at the Eddington limit corresponds to a $0.5-2$ keV flux of $\sim 10^{-17}$ erg s$^{-1}$ cm$^{-2}$; however, an instrument like Athena reaches the confusion limit at a few $\times 10^{-17}$ erg s$^{-1}$ cm$^{-2}$ \citep{Piro2023} due to its larger PSF (5\arcsec on-axis, and larger off-axis) which would fundamentally limit its ability to distinguish the counterpart. AXIS, on the other hand, has a much lower confusion limit due to its high angular resolution across the FOV, allowing the detection of faint post-merger emission.

Even if LISA is not operational at the same time as AXIS, AXIS will still contribute to LISA science in several important ways. First, the EM searches for the progenitors of MBH mergers (i.e., MBHBs with longer orbital periods) prior to LISA's launch could yield the massive and extragalactic analogs of ``verification binaries'' (which are known Galactic compact binary systems whose loud GW signals guarantee that LISA would detect them; e.g., \citep{Stroeer2006}) and assist the crucial tests of the instrument in the early phase of the mission. Furthermore, the AXIS survey fields and serendipitous observations of AGN from Guest Observer programs could provide a deep reference catalog ($\sim 2$ orders of magnitude deeper than the all-sky eROSITA) for future X-ray follow-up: \citet{Lops2023} showed that using an X-ray reference catalog would exclude a fraction of AGN as hosts within the LISA localization error box, thereby reducing the follow-up effort. Finally, X-ray observations of any MBHBs (and their mergers) with AXIS would help constrain expected detection rates, provide important test beds for studying their EM emission, and offer valuable lessons for devising follow-up strategies for LISA detections.

In addition to being astrophysically rich systems in their own right, multi-messenger observations of MBHBs also have significant implications for fundamental physics and cosmology. For instance, MBHBs can be used as standard sirens to probe the expansion of the universe out to high redshifts ($z\sim10$) through the luminosity distance-redshift relation (e.g., \citep{Tamanini2016}). In binary periodicity models where the EM emission is phase-linked to GWs (i.e. Doppler boosting and self-lensing), simultaneous observations of the EM and GW signals of the same MBHB source could even place constraints on the graviton mass and alternative theories of gravity, by comparing the propagation speed of GWs versus light \citep{Haiman2017,Davelaar2022}.

\section{Supernovae}
\label{SNe}

\subsection{Shock Breakout Emission}
\label{sec:SBO}
The most straightforward means of progenitor identification of core-collapse (CC) SNe (i.e., those with massive star progenitors) is via direct detection in pre-explosion imaging. Relying largely on Hubble, this technique has firmly established red supergiants (RSGs) as the progenitors of Type IIP SNe (H-rich spectra with long-lived light curve plateaus; \cite{Smartt2009}). For other subtypes of SN, direct progenitor associations are extremely rare: the handful of examples to date include a single blue supergiant (BSG) for SN 1987A \citep{Sonneborn1987, Walborn1987}, several yellow supergiants (YSG; \citep{VanDyk2014, Tartaglia2017, Kilpatrick2017}), and one stripped-envelope SN of uncertain progenitor type \citep{Folatelli2016}. But because SNe are so bright, inevitably most are simply too distant for their pre-explosion progenitors to be directly studied. Another technique is desperately needed for the majority of SNe at distances beyond which HST can detect progenitors ($d > 30$\,Mpc).

\begin{figure}[H]
    \centering
    \includegraphics[width=0.45\columnwidth]{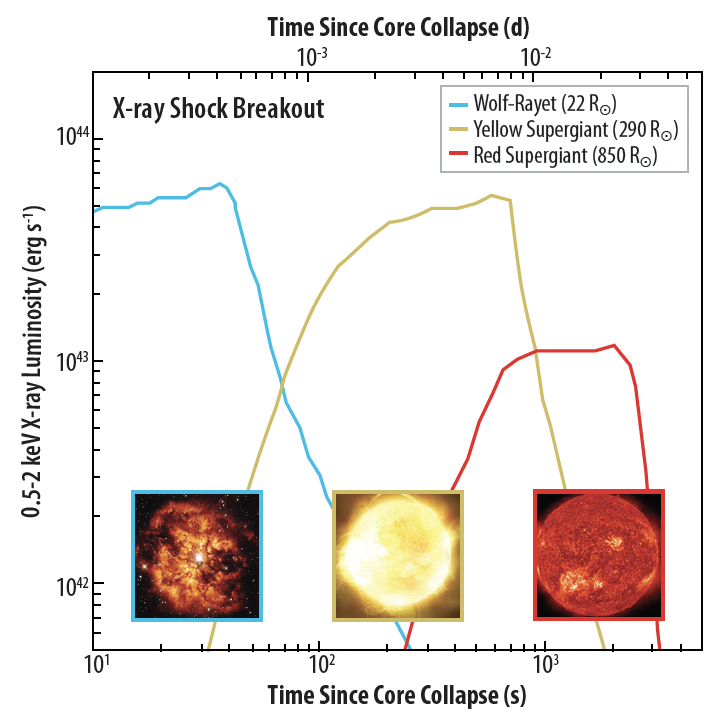}
    \includegraphics[width=0.45\columnwidth]
    {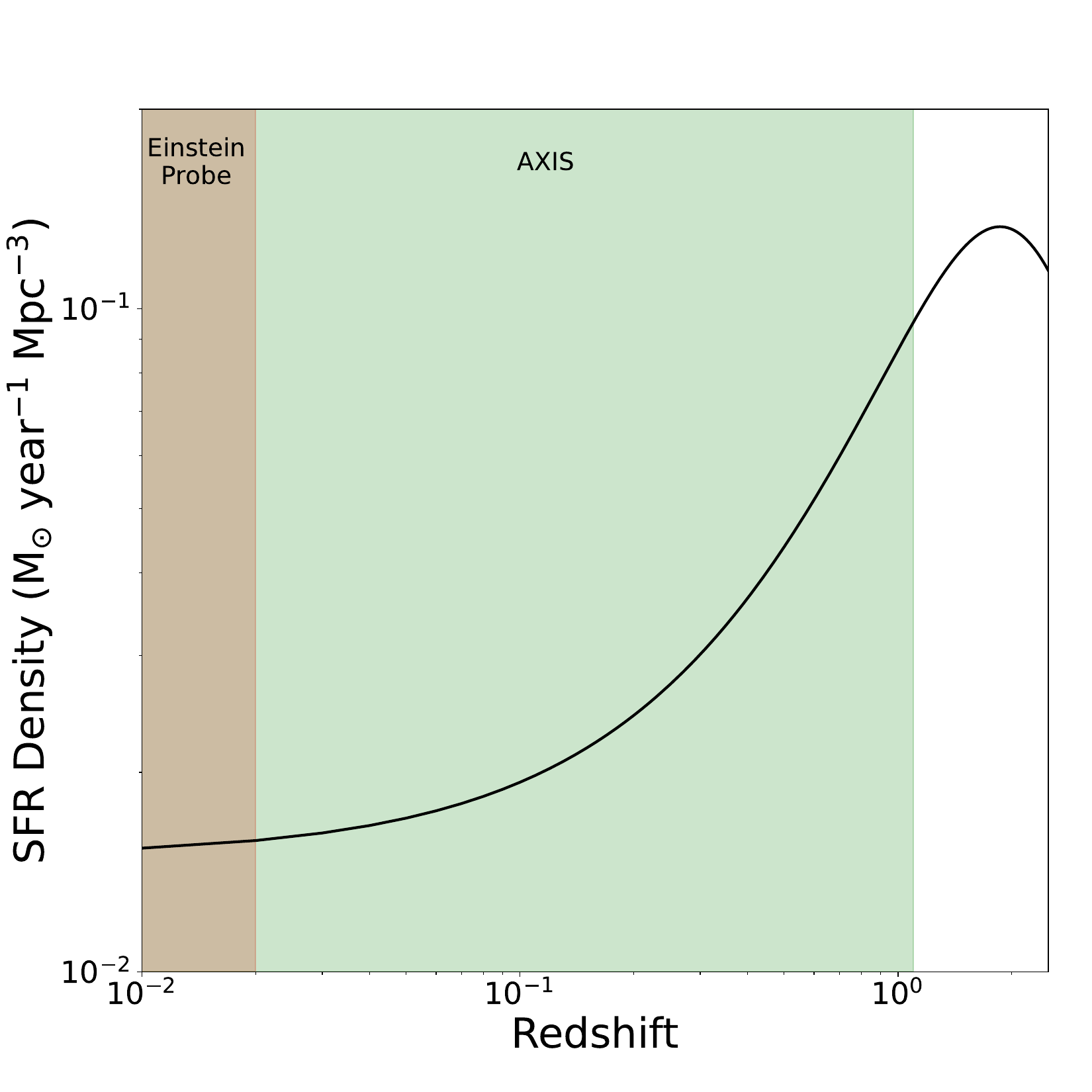}
    \caption[AXIS sensitivity to shock breakout emission from core-collapse supernovae]{AXIS will serendipitously discover shock breakout (SBO) emission from core-collapse SNe out to $z\approx1$, uniquely revealing fundamental properties of the progenitor star. Left: The duration (and spectrum) of the shock breakout signal is a strong function of the progenitor radius. With the large span of radii observed in massive stars, X-ray SBO discoveries can identify progenitor stars at distances far beyond direct progenitor imaging. Right: Unlike shallow/wide X-ray facilities such as Einstein Probe, AXIS will detect SBO near the peak of cosmic star formation, uniquely probing its evolution over cosmic time.}
    \label{fig:SN_SBO}
\end{figure}

AXIS will transform our view of the end states of massive stars through the serendipitous discovery of shock breakout (SBO) emission --- the moment when the first electromagnetic radiation escapes from the exploding star. The duration and spectrum of the SBO signal provide a direct measure of the radius of the progenitor (Figure~\ref{fig:SN_SBO}) –-- for example, compact progenitors such as Wolf-Rayet stars are expected to result in SBO signals of duration $\sim$ minutes with SEDs that are predicted to peak in the soft X-rays, while more extended objects such as red supergiants will have correspondingly longer and cooler prompt SBO signals \citep{Nakar2010, Waxman_FXT_2017}. Thus unlike wide-field UV and optical surveys, X-ray SBO discoveries can uniquely probe the progenitors of the stripped-envelope core-collapse supernovae (e.g., Type Ib/c), precisely the sources that are most poorly constrained from pre-explosion imaging.

Currently only a single well-established example of a real-time X-ray SBO detection is known: SN2008D, a type Ib (He-poor) supernova in the nearby ($d = 27$\,Mpc) NGC2770. While upcoming wide-field X-ray observatories such as the Einstein Probe will likely uncover more such nearby examples, the unprecedented sensitivity of AXIS will enable the detection of SN2008D-like events out to $z\sim \! 1$, near the peak of cosmic star formation \citep{Madau2014}. As a result, AXIS will measure the evolution of progenitor properties, as well as the stripped-envelope core-collapse supernova rate, in an entirely unique manner. These discoveries are entirely serendipitous and do not require dedicated observational time. And with its on-board transient detection algorithm, AXIS SBO discoveries will be rapidly downlinked and disseminated to the broader astronomical community, enabling prompt multi-wavelength follow-up to characterize the subsequent shock-cooling emission, as well as the radioactively powered phase.

\begin{figure}[H]
    \centering
    \includegraphics[width=0.8\columnwidth]{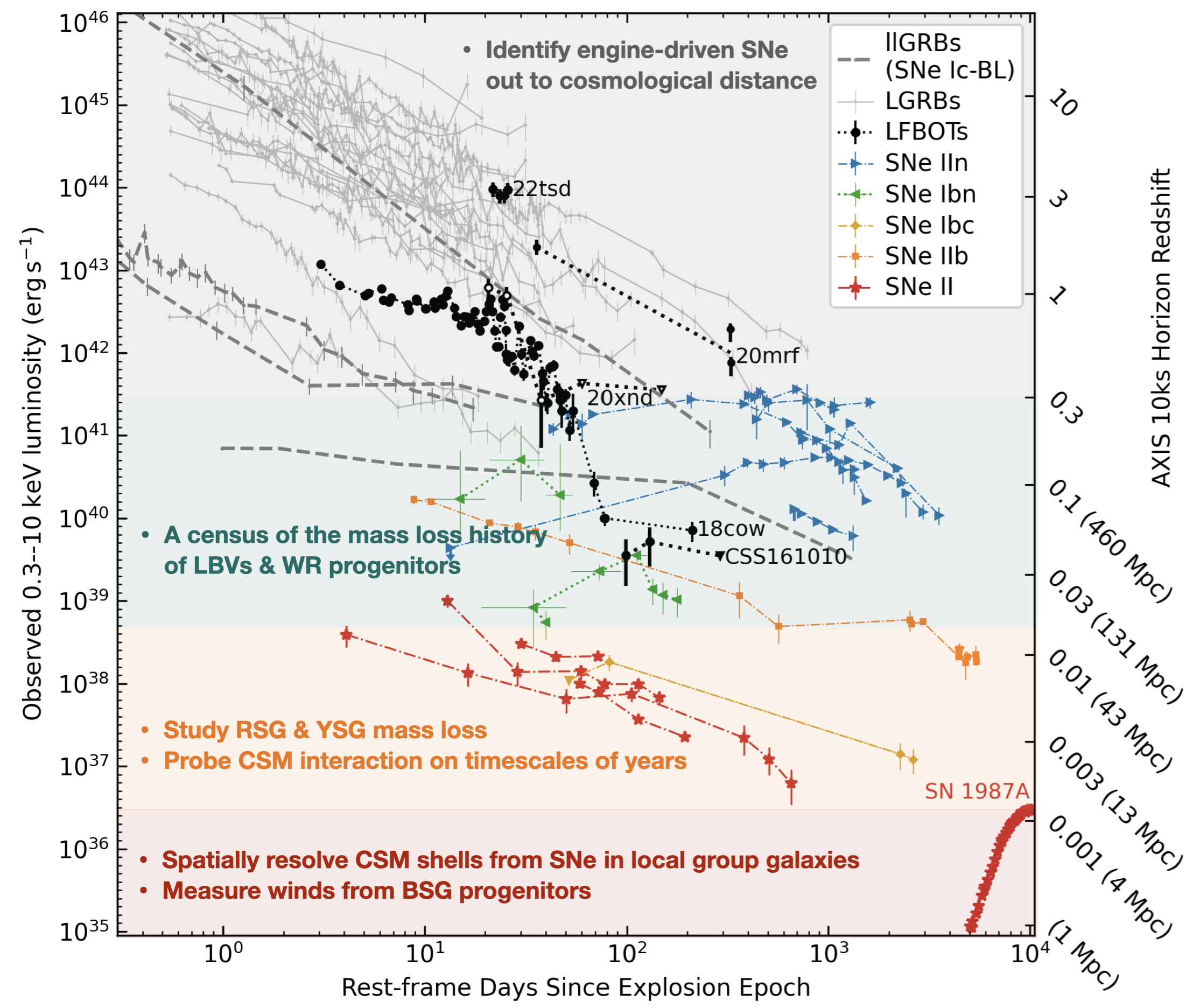}
    \caption
    [AXIS constrains core-collapse SNe spanning a wide range of X-ray luminosity]
    {AXIS offers a unique opportunity to study all types of core collapse SNe (CCSNe), spanning a wide range of X-ray luminosity. 
    Here we show massive star explosions with well-measured X-ray light curves (see \citep{Dwarkadas2012, Yao2022a} and references therein).
    The y-axis on the right side shows the AXIS horizon redshift. 
    \label{fig:SNe_xlc}
    }
\end{figure}

\subsection{Circumtellar Medium Diagnostics}
\label{sec:csm}
The formation of SNe is closely related to the end-of-life evolution of massive stars, which is a phase of stellar evolution that remains poorly constrained \citep{Smith2014}. When a star explodes as a SN, a shock wave driven by the explosion propagates into the circumstellar medium (CSM) shaped by prior mass loss of the progenitor. The particles behind the shock wave are accelerated to relativistic speeds, producing inverse Compton emission at early times ($\lesssim 30$\,days) and thermal bremsstrahlung with line emission afterwards (see, e.g., \citet{Chevalier2017, Chandra2018} for reviews). Since the shock velocities are much faster than the speed of the progenitor's mass ejection, the X-ray and radio emission produced by shock interactions during the first few years carries imprints of the mass loss history up to thousands of years prior to their stellar death. 

Among the handful of CCSNe with well-sampled X-ray light curves, the few SNe with narrow emission lines from dense CSM (type IIn) are the most well-studied \citep{Dwarkadas2012, Chandra2015, Dwarkadas2016, Katsuda2016, Brethauer2022}. The X-ray luminosity from some of them typically reaches a peak of $10^{40}$--$10^{41}$\,erg\,s$^{-1}$ at a few years after explosion (see Figure~\ref{fig:SNe_xlc}), suggesting a CSM environment of 1--10\,$M_\odot$ out to $10^{16}$--$10^{17}$\,cm from the progenitor, ejected 10--100 years before the explosion \citep{Margalit2022, Vargas2022}. Such elevated mass loss rates of $\sim\!0.1\,M_\odot\,{\rm yr^{-1}}$ are indicative of a very massive luminous blue variable (LBV) or a LBV-like star. A few other SNe IIn appear to exhibit rather steep X-ray luminosity declines \citep{Pooley2002, Chandra2005}, pointing to RSGs with enhanced mass loss rates of $\sim\! 10^{-4}\,M_\odot\,{\rm yr^{-1}}$, likely due to nuclear burning instabilities or interaction within a binary system. In particular, the population of hydrogen-poor interacting SNe has been proposed to originate from compact Wolf-Rayet (WR) stars \citep{Foley2007, Perley2022, Gal-Yam2022}. However, existing X-ray observations are limited, encompassing merely two SNe Ibn (\citep{Immler2008, Ofek2013}; see Figure~\ref{fig:SNe_xlc}) and zero SNe Icn. By sampling the shock interaction emission of a representative sample of SNe Ibn and Icn, AXIS holds the promise for greatly advancing our understanding of the mass-loss history of WRs and LBVs during their transition to a WR-like state.

For the majority of type II SNe, progenitor stars have much lower mass loss rates of $10^{-6}$--$10^{-5}\,M_\odot\,{\rm yr^{-1}}$, and therefore have relatively under-luminous ($\lesssim 10^{39}\,{\rm erg\,s^{-1}}$) X-ray emission \citep{Pooley2002, Misra2007, Chakraborti2012, Chakraborti2016}. AXIS monitoring of nearby SNe II will provide critical insights into the mass loss from RSGs (and BSGs) both as a function of progenitor mass and lookback time before the explosion. Finally, the spatial resolution of AXIS offers a unique opportunity to study CSM structures in any extremely nearby SN, the probability of which is small but not negligible over the next 20 years [Astro 2020 Decadal Survey, B-DA6]. For example, Chandra observations of SN\,1987A over 16 years revealed an equatorial ring structure \citep{Frank2016}. Detailed studies of non-spherical CSM structures conducted by AXIS could unravel the CSM geometries encompassing CCSNe within galaxies of the Local Group or our own Milky Way.


\subsection{Compact Object Formation}
A long-standing observational challenge in SNe studies is the identification of the newly formed compact object (i.e., a neutron star or a black hole). In the majority of stellar explosions, the compact object stays either inactive or deeply embedded in the SN ejecta. The opportunity to investigate compact object formation arises primarily in a small fraction of ``engine-driven'' SNe, where the compact object consumes stellar materials, generates heat, and ejects outflows. 
For a collimated relativistic outflow pointing towards the observer, internal energy dissipation with the jet gives rise to a long-duration gamma-ray burst (LGRB; see \citet{Zhang2018} for a recent review), and the subsequent afterglow smoothly decays as $L_{\rm X} \propto t^{-1}$ in the X-ray band. The population of low-luminosity long GRBs (llGRBs) are found to be associated with type Ic broad line SNe \cite{Woosley+2006}.

This frontier of massive star deaths, thanks to optical time-domain surveys, continues to be vigorous. 
A recent observational breakthrough is the recognition of an emerging new class of engine-driven stellar explosions with sub-relativistic outflows, suppressed $\gamma$-ray prompt emission, and \textit{ long-lived} engine activities in the X-ray band \citep{RiveraSandoval2018, Margutti2019, Coppejans2020, Ho2020, Pasham2021, Bright2022, Ho2022, Yao2022a, Ho2022_tsd_discovery}. Existing research highlights six known events of this phenomenon, collectively referred to as luminous fast blue optical transients (LFBOTs) or AT2018cow-like events. The optical characteristics of these events are shaped by their remarkable energy release, low ejecta masses, and high temperatures. Five LFBOTs have existing X-ray observations (see the black circles in Figure~\ref{fig:SNe_xlc}), where the extremely luminous and rapidly variable X-ray emission from the prototype AT2018cow \citep{Margutti2019}, AT2020mrf \citep{Yao2022a}, and AT2022tsd \citep{Matthews2023} show compelling evidence for the presence of a central engine.  

With a sample of six events, the nature of LFBOTs remains a subject of active debate \citep{Perley2019, Kremer2021, Metzger2022_FBOT}. An expanded sample size, coupled with multi-wavelength observations spanning various evolutionary stages, holds the key to unraveling the connections between LFBOTs, llGRBs, and other SN types. 
Next-generation time domain surveys such as the Vera Rubin Observatory \cite{2019ApJ...873..111I} will extend the detection of such events from $z\sim 0.2$ to $z\sim1$. 
The sensitivity and rapid response of AXIS will be crucial to provide early-time X-ray observations to differentiate them from ordinary FBOTs (which are normal stripped-envelope SNe in dense CSM), and trigger rapid follow-up spectroscopy of the UV/optical/IR thermal emission (which only lasts for one month). 

Another fundamental open question in LFBOTs is the nature of their central engine. The key diagnostic is the decay rate of the X-ray light curve at $t\gtrsim 20$\,days, at which point the energy released from the central region transitions from being partially obscured to becoming mostly exposed \citep{Margutti2019}. If the engine is a rapidly spinning magnetar, it will deposit energy into the ejecta at a rate of $L_{\rm engine} \propto t^{-2.4}$ \citep{Metzger2018}, whereas fall-back accretion onto an accreting black hole gives $L_{\rm engine} \propto t^{-5/3}$ \cite{Rees1988, Phinney1989}. A late-time plateau phase or shallower decay will indicate the formation of an accretion disk. We note that the volumetric rate of LFBOTs is extremely small --- only 0.1--$0.01$\% of the CCSNe rate \citep{Ho2023}, or 10--100\,$\rm Gpc^{-3}\,yr^{-1}$ \citep{Perley2020}. Therefore, to construct a decent sample of LFBOTs with late-time X-ray measurements down to $\approx 10^{40}\,{\rm erg\,s^{-1}}$ (as shown in AT2018cow and CSS161010, see Figure~\ref{fig:SNe_xlc}), a horizon distance of a few hundred Mpc is needed, which speaks for the requirement of a sensitive X-ray instrument like AXIS.

\section{Tidal Disruption Events}\label{TDE}
A star coming too close to a massive black hole (MBH) gets disrupted by the tidal forces of the MBH around the tidal radius $R_{\rm T}$. Following this encounter, the debris evolves into an elongated stream, half of which comes back to form an accretion disk, producing thermal radiation that peaks in the EUV and soft X-ray \citep{Ulmer1999}. This emission can be reprocessed into the optical/IR bands by TDE debris, dusty torii, and cold gas. Assuming a flat distribution of the specific orbital energy, the debris fall-back rate ($\dot M_{\rm fb}$) initially rises for about one month, and then declines as $t^{-5/3}$ \citep{Rees1988, Phinney1989}. Over the past three decades, tidal disruption events (TDEs) have gone from theoretical curiosities to established transient phenomena \citep{Gezari2021}. As of 2023, $\sim \! 150$ TDEs have been reported. Among them, four objects are associated with on-axis collimated relativistic jets (known as ``jetted TDEs''), manifested by their extremely bright X-ray and radio emission (see \citet{DeColle2020} for a review). 

\begin{figure}[H]
    \centering
    \includegraphics[width=0.65\columnwidth]{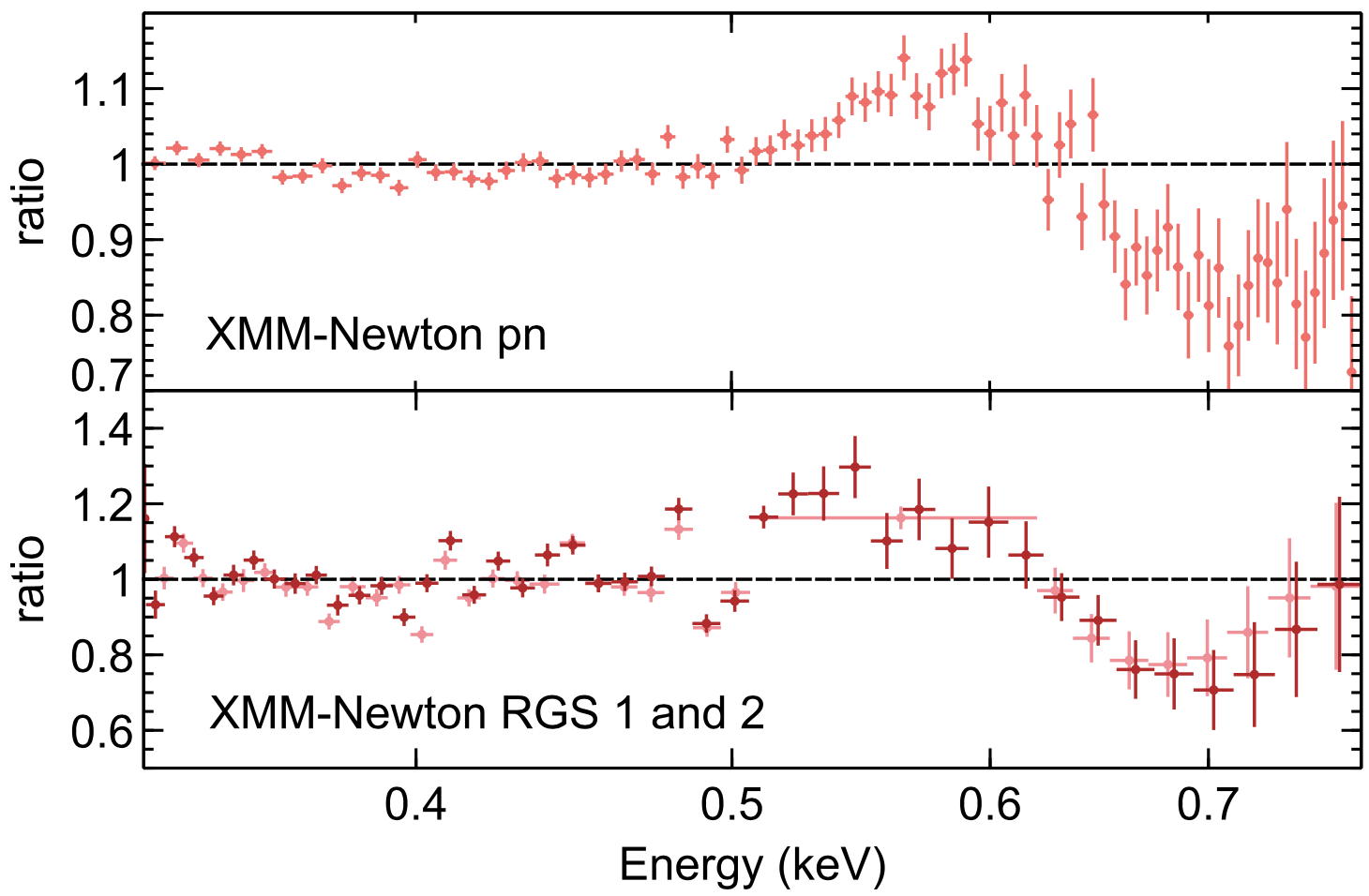}
    \caption[Ultra-fast outflows from tidal disruption event ASASSN-14li]
    {Ultra-fast outflows (UFOs) from the tidal disruption event ASASSN-14li will allow AXIS to probe super-Eddington accretion onto distant massive black holes. The ratio of the early-time XMM-Newton spectra of ASASSN-14li to a single \texttt{diskbb} model. The P Cygni-like absorption feature around 0.7\,keV is seen in both CCD and grating spectra. Figure adapted from \citet{Kara2018}. \label{fig:tde_ufo}}
\end{figure}

Since the peak TDE mass fall-back rate is above the Eddington limit ($\dot M_{\rm fb, peak}\sim\! 10^2(M_{\rm BH}/10^6\,M_\odot)^{-3/2} \dot M_{\rm Edd}$), TDEs provide unique laboratories to study super-Eddington accretion, the physics of which is highly uncertain. For example, fast outflows are often produced in numerical simulations as a result of the high radiation pressure \citep{Ohsuga2011, Jiang2014, Sadowski2016, Curd2019}. However, observationally such ultra-fast outflows (UFOs) are poorly characterized. In the X-ray, UFOs manifest themselves as blue-shifted absorption lines on top of the continuum emission. Among the known TDEs, such features have been suggested in a handful of TDEs, e.g. ASASSN-14li  (Figure~\ref{fig:tde_ufo}; \citep{Kara2018}). AXIS spectroscopic monitoring campaigns of nearby TDEs will enable us to systematically study the evolution of the outflow velocities and energetics. 

TDEs provide an ideal testbed for studying the physics of accretion through all regimes. As the mass fallback rate decreases, a natural prediction is that the disk may undergo a thermal-viscous instability caused by the change in advective heat transport and radiation pressure, triggering a state transition from a radiation pressure dominated thick disk to a gas pressure dominated thin state \citep{Kato2008, tchekhovskoy14_mad_jet, Shen2014, Lu2022_tde_review}. The latter geometrically thin disk lacks the capacity to confine the horizon-threading magnetic fields, which results in striking variations in inflow and outflow dynamics, such as the shutoff of relativistic jets and the destruction of a magnetically dominated corona. For example, by capturing the sudden reduction in X-ray luminosity and variations in the X-ray spectral shape, state transitions have been identified in two jetted TDEs (Swift\,J1644+57 \citep{Zauderer2013} and Swift\,J2058+05 \citep{Pasham2015}) and two non-jetted TDEs (AT2018fyk \citep{Wevers2021} and AT2021ehb \citep{Yao2022b}). However, X-ray monitoring of ASASSN-14li for 500\,days shows no evidence for a sudden change in luminosity, which could indicate a disk instability or state transition \citep{Brown2017}. Since the luminosity of the thin disk state ($\sim \!10^{41}$\,erg\,s$^{-1}$) is generally much fainter than that of the thick disk state ($\sim \!10^{43}$\,erg\,s$^{-1}$), a complete characterization of TDE state transitions requires sufficient X-ray sensitivity. Future long-term AXIS monitoring of nearby TDEs (selected by optical or X-ray surveys) is needed to reveal the prevalence, timescale, and physical conditions of thermal-viscous instability in accreting MBHs.

X-ray quasi-periodic oscillations (QPOs) have been detected from Swift J1644+57 and ASASSN-14li \cite{Reis2012,Pasham2019}. The short timescales (mHz) are consistent with an origin
in the innermost accretion flow. These QPOs provide a means to study the relativistic spacetime and constrain the black hole spin in this population of hitherto dormant black holes. Current searches with Chandra and XMM-Newton are limited to the most luminous TDEs. The occurrence rate of this quasi-periodic variability, and its relation to the evolution of the broader TDE accretion flow, is uncertain. AXIS will facilitate detection of these signals in a larger sample of less luminous TDEs.

\section{Changing Look/State AGN}
This subclass of unbeamed AGN splits into two relatively well-defined groups, both of which exhibit dramatic variations in their X-ray, UV, optical continua and/or broad emission lines on timescales of days to years (see \citet{Ricci22} for a recent review). One subset, known as Changing-Look (CLAGN; term coined in \citet{Matt2003}) or Changing-Obscuration AGN (COAGN), show strong line-of-sight column density changes (most dramatically from Compton thin to thick or vice versa), mostly associated with clouds or outflows eclipsing the central engine of the AGN. The other subset, known as Changing-State (CSAGN) or rather confusingly ("optical") CLAGN, shows continuum and broad emission lines which appear or disappear, typically triggered by strong changes in the accretion rate of the SMBHs.

Over the past few decades, our understanding of CLAGN/COAGN has improved substantially, thanks to the monitoring campaigns of many local AGN. For example, NGC 1365 shows strong X-ray obscuration transitions ($N_{H}{\sim}10^{22}$-$10^{24}$ cm$^{-2}$) on time-scales of weeks \citep{Risaliti2005}, days \citep{Risaliti2007}, and even $\sim$10 hours \citep{Risaliti2009}, implying cloud densities as high as ${\sim}10^{11}$ cm$^{-3}$ and distances of 3000-10000\,$r_{g}$, with densities consistent with those expected in the Broad Line Region (BLR). In rare cases with exceptional statistics, the complex geometry of the clouds has been probed, suggesting comet-like structure \citep{Maiolino2010}. Ensemble studies of several dozen other local AGN find transient obscuring clouds at similar radii, which implies that the clouds are generally just inside or near the dust sublimation radii for these AGN \citep{Markowitz2014}. Finally, even Compton-thick AGN like NGC~1068 exhibit such eclipses \citep{Zaino2020}, highlighting that clumpy variable obscuration from the BLR or the torus is a very common property among all types of AGN. 

However, many open questions regarding CLAGN/COAGN that can provide insight into the torus and BLR structure remain. For example, how does the rate and distribution of obscuration events (and perhaps even the individual cloud properties) relate to the line-of-sight orientation, optical AGN type, state of the torus and BL region clouds, and host galaxy properties? What fraction of AGN show changes in obscuration and why? To answer such questions, we need to greatly increase the number of objects and the fidelity of the constraints on the $N_{H}$ variations with time. AXIS will do just this, greatly expanding the statistics to many thousands of relatively bright AGN, allowing studies along across nearly every potential vector, pinning down the cloud properties across much larger ranges of black hole mass, accretion rate, and obscuration/AGN type parameter spaces, to understand cloud occurrence rates, sizes and shapes, and ensemble distributions.

The vast majority of CSAGNs discovered to date were found using archival data, leaving a full account of their behavior largely unexplored. One of the first CSAGN to be extensively tracked was 1ES\,1927+654 \citep{Trakhtenbrot2019,Ricci2020,Laha+2022}, which exhibited X-ray/UV/optical luminosity variations by a factor of $>$50 on few-month timescales. This CSAGN highlights the potential for variations in both the optical and X-ray regimes, and was extensively monitored at X-rays and optical through its transition state, allowing strong constraints on the physics of the system. Its origin was argued to be a TDE, which provoked an increase in the accretion rate at the innermost regions of the accretion disk, which then emptied the inner disk and led to the destruction of the X-ray corona. 

More generally, such objects can provide valuable insights into the dynamic processes occurring in the immediate vicinity of supermassive black holes, how matter behaves under extreme gravitational forces, and how AGN evolve over time and affect their host galaxies. However, the origin of the anomalous accretion disk variability is not yet well understood. The overall demography of CSAGN is also poorly known, due to the lack of dedicated multi-wavelength time-domain surveys and/or their relative sensitivities. For example, \citet{Temple2023} derive a CSAGN rate of 0.7--6.2\% on 10--25 yr time-scales among local Swift-BAT selected AGN  objects (i.e., 21 of 412), with many transitions occurring within at most a few years and nearly all having ${<}0.1L/L_{\it Edd}$.

There are also many open questions here. From an X-ray standpoint, how common are Changing-State transitions among AGN as a function of their fundamental properties ($M_{\rm SMBH}$, accretion rate, spin, radio-loudness), what are the typical changes in the X-ray emission during the transition, what are their typical transition timescales, and whether they (and how often) they repeat.
Large samples of CSAGN, together with high-cadence monitoring, are required to understand their occurrence rates and typical timescales.
AXIS can revolutionize our understanding of CSAGN. Its enhanced sensitivity will allow the detection of fainter and more distant sources, expanding the sample size and improving our understanding of their demographics, providing better photon statistics to probe spectral changes, yielding insights on accretion physics and the dynamics of circumnuclear regions.

\section{Quasi-Periodic Eruptions}
\label{sec:qpe}
X-ray quasi-periodic eruptions (QPEs) are a new phenomenon that was recently discovered from a handful of low-mass galaxies (with stellar masses $\approx 10^{9-9.5} M_{\odot}$) in the nearby (within $z\sim0.05$) Universe \citep{miniutti19,giustini20,arcodia21}. They are sharp soft X-ray bursts that last from less than an hour to a few hours and repeat in a quasi-periodic fashion every several hours to almost a day, although some sources show a large scatter in the recurrence \citep{giustini20,arcodia21,arcodia22}. Their origin is consistent with the nuclei of their host galaxies and most of the sources are detected in X-rays between the QPEs, with a spectral shape consistent with the exponential decay of a thermal spectrum (with peak temperature $kT \sim40-80\,$eV, \citep{miniutti19,giustini20,arcodia21}). In similarity with TDEs, this quiescence spectrum is interpreted as emission from the innermost region of the accretion flow, indicative of a black hole with a rather small mass. Estimates from scaling relations confirm that QPEs originate around massive black holes of $M_{BH}\sim10^{5-6.7} M_{\odot}$ \citep{Wevers+2022:qpehosts}. When in eruption, the increase in the soft X-ray count rate is usually a factor of 10-100, and the spectrum remains soft, following a characteristic spectral evolution showing a harder rise than decay at the same count rate \citep{arcodia22,miniutti23}. 

At the time of writing, only four publicly known extragalactic repeating soft X-ray erupters show this characteristic behavior and are, therefore, considered secure QPE sources \citep{miniutti19,giustini20,arcodia21}. Two further candidates show a similar energy dependence during the bursts, but the observations were not long enough to constrain the possible repetition \citep{Chakraborty+2021:qpecand,Quintin+2023:qpecand}. So far, no simultaneous variability has been observed in other bands (e.g., optical, UV, IR or radio), although this might be due to the angular resolution of current observations, which are most likely dominated by the galaxy’s stellar population.

The origin of QPEs is still debated. Since their discovery in 2019, several models have been proposed, including accretion disk instabilities \citep[e.g.,][]{Sniegowska+2020:cl,Sniegowska+2023:instab,Kaur+2023:instab,Pan+2023:instab} or scenarios involving a two-body system consisting of a massive black hole and a much smaller companion \citep[e.g.,][]{King2020,king22,Sukova+2021:coll,Xian+2021:qpes,Zhao+2022:qpes,Wang+2022:qpes,Metzger+2022:qpes,Krolik+2022:qpes,Lu+2023qpes,Linial+2023:qpes,Linial+2023:unstable,franchini23,Tagawa+2023:qpes}. The most recent models within the latter scenario seem to reproduce the observational properties at least qualitatively \citep{Lu+2023qpes,Linial+2023:qpes,franchini23,Tagawa+2023:qpes} and propose QPEs to be triggered as a star or black hole brought into the nucleus via extreme mass ratio inspirals (EMRIs) passes through the accretion disk around the massive black hole. This accretion flow may be provided by a previous TDE, which would be supported by the growing connection between QPEs and the signature of potential previous TDEs \citep{miniutti23,Sheng+2021:tdegsn,Chakraborty+2021:qpecand,Quintin+2023:qpecand}. This scenario, if confirmed, would make QPEs the first electromagnetic counterpart of EMRIs, opening a new window to the future of multi-messenger astronomy.

AXIS will perform follow-up of QPEs discovered by either AXIS itself or other X-ray surveys, with the aim of characterizing their short- and long-term variability and their spectral evolution. Long and continuous observations with a sensitive soft X-ray imager (most of the signal from QPEs is below $1.5-2$\,keV) are needed, and AXIS will play a central role on the study of QPEs. Furthermore, since in quiescence these sources are often detected with a soft thermal component, similar to that of TDEs, indicative of an accretion disk, AXIS spectra will be used to put constraints on the mass and spin of the massive black hole.

\section{Magnetars}
\label{sec:magnetars}
Magnetars are highly magnetized, rotating young neutron stars that display a wide range of radiative activity (see, e.g., \cite{kaspi2017magnetars} for a review). The most energetic bursting phenomena are called magnetar giant flares (MGF), with a peak luminosity of $10^{41}-10^{47}$ erg s$^{-1}$. Such flares are associated with non-disrupting powerful explosions breaking through the neutron star surface and ejecting a relativistic and collimated outflow. Only a handful of such events have been observed so far from both Galactic and extragalactic magnetars. They were characterized by a very short ($\lesssim$100\,ms), energetic ($10^{44}-10^{46}$\,erg), hard spike immediately followed by a minutes-long decaying tail modulated by the spin period of the magnetar (usually on the order of seconds). Both signatures are observed for all nearby events ($\lesssim$0.5\,Mpc), but in more distant events, only the prompt spike has been detected with current instruments (e.g., \cite{2021ApJ...907L..28B}). The characteristic periodic tail is the smoking gun signature for the association of such short transients to MGF. Generally, the phase-integrated spectrum of the tail is well described by a thermal blackbody component peaking around 5 keV plus a non-thermal power-law component that emerges above tens of keV. The spectrum of MGF tails evolves in time, with a thermal component temperature moving from $\sim 10$ keV to $\sim 3$ keV, and the study of the temporal evolution revealed the presence of quasi-periodic oscillations (QPOs) with typical range of frequencies between 10 and 1000 Hz \cite{Israel2005,Strohmayer2005,Watts2006}. QPOs are associated to oscillations of the stellar crust or inner layers of the star. They can inform us about the structure and properties of the dense matter that constitutes neutron stars, e.g., its equation of state, and give us insights on the crust-core interface, thought to be a crucial link to the physical origin of the flares \cite{Watts+2007,Huppenkothen+2014}.


A recent population study inferred a volumetric rate of MGFs oriented towards Earth of $R_{MGF}^{Vol}=3.8_{-3.1}^{+4.0}\times10^5$\,Gpc$^{-3}$\,yr$^{-1}$ \citep{2021ApJ...907L..28B}. Such a high volumetric rate, together with the small sample of observed events to date, highlights the need of a sensitive soft X-ray telescope that could detect and resolve the smoking-gun signature of the periodic tails of MGFs. Observations can be triggered by hard X-ray monitors, which would also provide localization. A fast response to such a trigger is essential. Once on target, the flux sensitivity will be key to detect the fading tail and reveal the presence of periodic emission. Unambiguously identifying extragalactic MGFs is imperative to learn more about magnetars as a population and, in general, to provide more information for population synthesis models.

Some Galactic magnetars show longer-lived X-ray flux enhancements (factor of 10-1000), lasting for weeks to months. When this happens, it is said that the magnetar is in ``outburst''. Anomalies such as spectral hardening, glitches, changes in the pulse profile, and repeated shorter bursts can occur during these time periods \citep{kaspi2017magnetars}. The number of outbursts per magnetars can vary from none to several in a decade \citep[see, e.g.,][]{Guver_2007, 2007ApJ...654..470W, 2011ApJ...729..131N}. A strategic monitoring of outbursting magnetars by a sensitive X-ray telescope is crucial to uncover the mechanisms in place, ultimately giving clues on the physics and internal structure of these neutron stars. For example, flaring SGRs have recently been identified as a source of (at least a sub-population of) fast radio bursts (FRBs) \cite{Bochenek+20,Ridnaia+21}. This also remains the only instance of an FRB detection coincident with X-rays. We discuss in Section~\ref{sec:FRB} the significance of such multiwavelength detections to anchor models of magnetars, and FRB emission mechanisms on a solid footing.

When short bursts and flares are produced by Galactic magnetars, the emitted soft X-rays interact with surrounding dust in the Milky Way. Such dust is often structured in dense clouds along the line-of-sight and creates dust-scattering haloes or rings (one for each dust cloud). This phenomenon has been observed several times \citep[see, e.g.,][]{2010ApJ...710..227T}, highlighting the crucial aspects of the sensitivity and angular resolution of the X-ray observations. Having a point spread function narrower than the rings' radial profiles enables the study of the dust clouds' thicknesses and composition, providing a unique observable to trace and characterize the dust in our Galaxy.

On top of their bursting activity, magnetars are fascinating X-ray persistent emitters (additional discussion about the potential of AXIS on this topic can be found in the accompanying AXIS White Paper ``Prospects for Compact Objects and Supernova Remnants Studies with AXIS''). Furthermore, a wind nebula has been observed surrounding the magnetar Swift~1834.9-0846 \citep{Younes_2016}. A more sensitive X-ray observatory could lead to the discovery of fainter magnetar wind nebulae, unveiling the connection between pulsars and magnetars, and providing more clues to the progenitors of such highly magnetized neutron stars.

\section{X-ray Binaries}
\label{sec:XRBs}
Observations of X-ray binaries (XRBs) probe black holes (with relevance to understanding super-massive black hole evolution via the scale-invariant nature of accretion), neutron stars (dense matter equation of state, high $B$-fields), and white dwarfs (stellar populations, LISA sources) with studies of all of these sources providing opportunities to constrain the theory of General Relativity. The integrated luminosity of high-mass X-ray binaries (HMXBs) is known to be correlated with the SFR in galaxies
\cite{2003_grimm_hmxb_sfr,2013_fragos_hmxb_sfr_sim}, while the integrated luminosity of low-mass X-ray binaries (LMXBs) is observed to correlate with the galaxy stellar mass \cite{2004_gilfanov_lmxb_mstar}, and thus their study is required to inform models for galaxy formation and evolution. Additionally, XRBs may play a key role in the re-ionization of the Universe at early times (e.g., \cite{2011_mirabel_xrb_reion,2013_fragos_xrb_reion,2023_liu_xrb_reion}).

HMXBs are highly variable and their study in the MW/LMC/SMC has revealed evidence of
variations at the population level in response to metallicity and age (e.g., \cite{2010_smc_hmxb_age,2021_lehmer_hmxbz}). Studies of these
systems with AXIS will provide insight into massive star formation in conditions akin to
those more typical in higher-$z$ galaxies \cite{2022_eldridge_araa,2023_fornasini_hmxbs_review}. In an HMXB, the X-rays from the compact object probe the wind of the massive star, providing detailed constraints on mass loss from massive stars, a key input to understanding their impact on star and galaxy formation \cite{nunez2017,2022_eldridge_araa}. 

Observations of the ULX population at larger distances will enable a detailed spectral characterization of the large and highly variable ULX population discovered by XMM-Newton and Chandra and inform our understanding of accretion at Eddington and super-Eddington rates \cite{2023_king_ulx_newar,2022_walton_ulxcat,2022_barrows_ulxz,2019_barrows_hlxcat}. 
AXIS spatial resolution is required to further the study of ULX environments (e.g., globular clusters \cite{Dage2020,Usher23}), complementing the Rubin Observatory \cite{Usher23}. 
The study of ULXs can
provide direct constraints on massive star evolution and LIGO binary merger progenitor
channels
\cite{2022_belczynski_uncertainty,2020A&A_belczynski_ligo_pop}. 

\begin{figure}[t!]
\centering
\includegraphics[width=0.49\columnwidth]{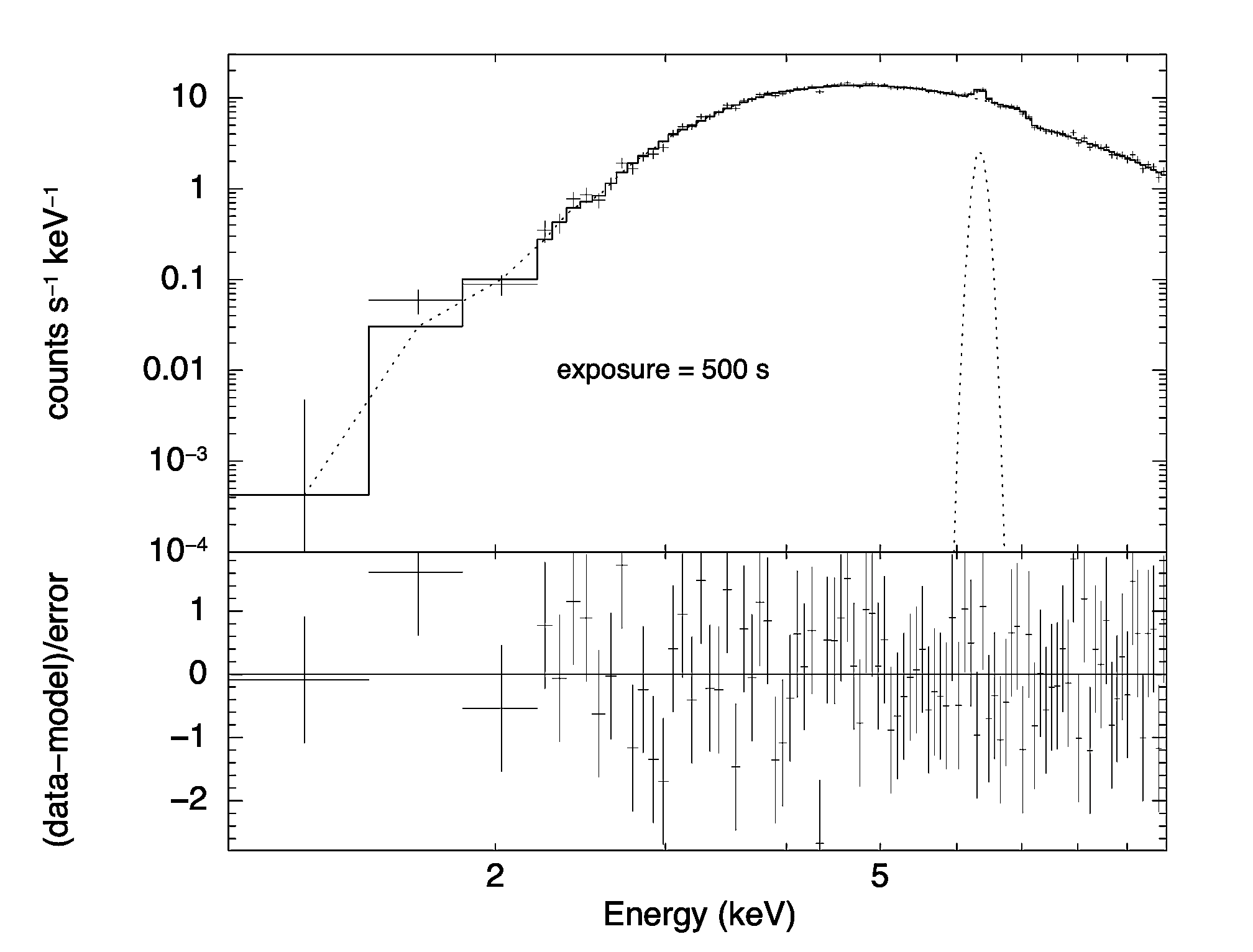}
\includegraphics[width=0.49\columnwidth]{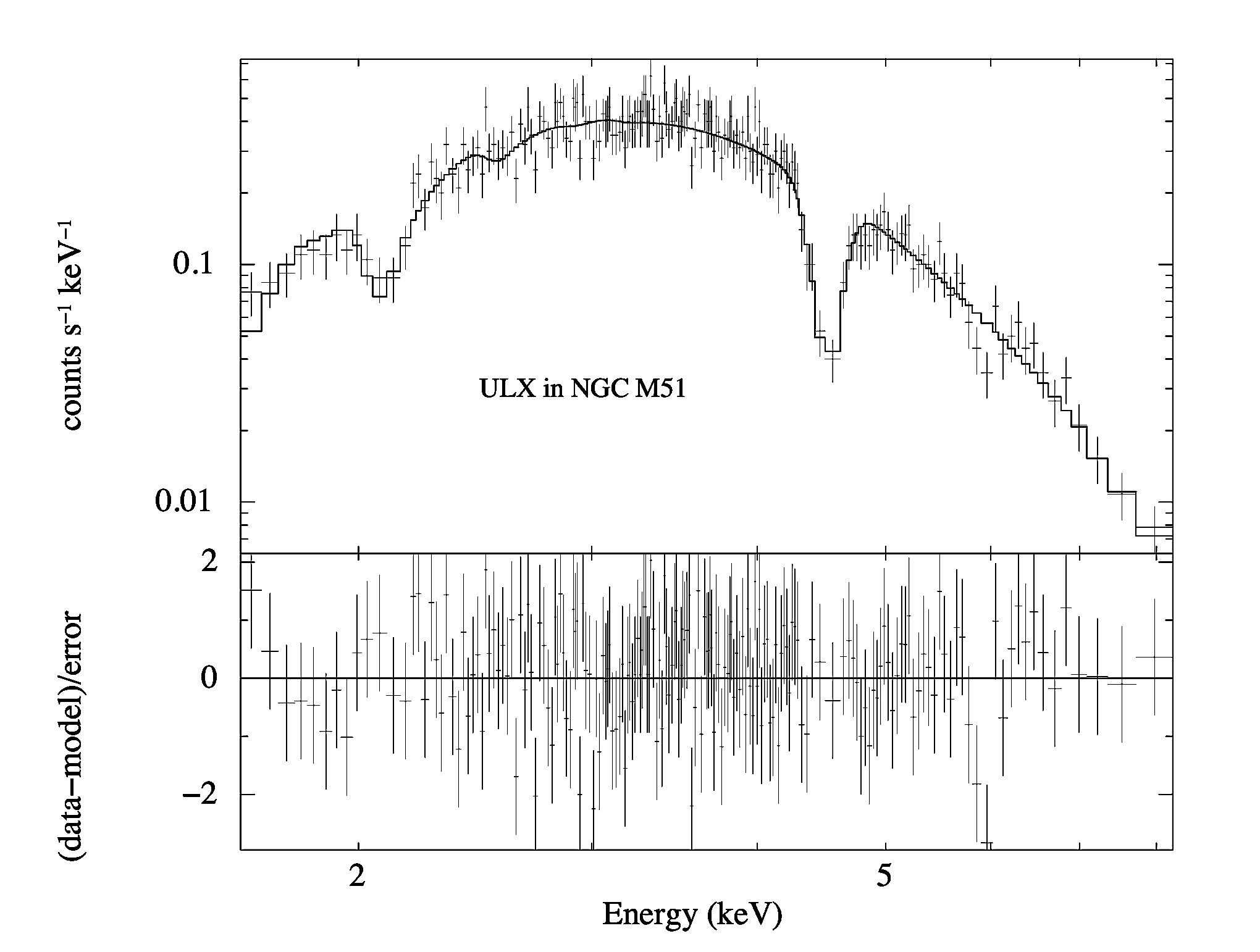}
\caption[Simulated AXIS spectra of Vela X-1 and ULX-8 in M51.]
{Left: Simulated AXIS 500\,s spectrum of Vela X-1. Right: Simulated AXIS 5\,ks spectrum of ULX-8 in M51, depicting the reported absorption feature, likely a cyclotron line \cite{brightman2018N}. The capability of AXIS to constrain the spectral parameters of XRBs in such short exposures will permit the study of fast time variability in exquisite detail.}
\label{ulx}
\end{figure}

Another possibility will be the search for cyclotron lines from the spectrum of ULXs. Recently, there has been at least one detection of a cyclotron line at 4.5\,keV \cite{brightman2018N}. This absorption can be tied to either electron or proton transitions, both of which result in different magnetic field strength estimates. Electrons imply $B = 4(1 + z) \times 10^{11}$ G for a 4.5 keV line, while protons suggest $7(1 + z) \times 10^{14}$ G. The observed line's unique broadening ratio of 0.02 aligns with a proton cyclotron resonant scattering feature (electrons yield broader lines), akin to theoretical predictions. We show in Fig.~\ref{ulx}, the power of the large effective area of AXIS to make such detections in exposures as short as 5\,ks.

%
%

AXIS will discover many XRBs in the local group
(MW/LMC/SMC/M31/M33) and beyond. The flexibility of scheduling and the high time resolution of AXIS will enable the study of transient and variable XRBs, which provide access to accretion flows that span
$10^{-9}~L_{Edd} - L_{Edd}$ in a single system on timescales of weeks to months, with dramatic
changes to the physical properties of the accretion flow occurring on timescales as short as the dynamical timescale
at the innermost regions of the accretion flow ($t_{dym} \sim$ kHz; \cite{2019_kara_maxij1820, 2023_ma_maxij1820}). Tracking the spectral and temporal characteristics as functions of luminosity is essential to understanding modes of accretion, the structure of the accretion flow, and the compact object. AXIS will provide new insight into accretion flow structure via studying low luminosity accretion flows ($L_x \lesssim 10^{-3}~L_{Edd}$). It will obtain high-quality observational constraints on radiatively inefficient accretion flow/jet physics (ADAFs, RIAFs) and constrain how accretion flows evolve towards/away from the high-luminosity thin-disk regime \cite{2017_plotkin_v404cyg,2013_plotkin_qbh,2013_qiao_adaf}. These constraints can be directly compared to the
latest generation of MAD/SANE numerical accretion flow models \cite{2021_kimura_mad_sim,2014_yuan_adaf_araa}. 

AXIS will constrain uncertain X-ray source populations, such as low luminosity X-ray transients discovered in Chandra observations along the Galactic plane \cite{2015wijnands_faint_lmxbs,2013_armasP_vFXTs}. Multi-epoch observations by AXIS will probe the nature of uncharacterized X-ray sources by constraining their spectral and flux variability, e.g., the nuclear cluster of stellar-mass black hole candidates \cite{2018_Hailey}. 
At the shortest orbital periods, a subset
of these systems will provide ``bright'' GW sources for LISA, requiring detailed electromagnetic characterization \cite{2004_nelemans_amcvn_gw}.

AXIS will detect and monitor the motion/decay of relativistic X-ray ejections from stellar
mass black holes in unprecedented detail, extending the baseline over which they can be
studied and providing insight into how the energy from the relativistic jet is dissipated through
interaction with the ISM
\cite{2002_corbel_xtej1550_jets,2017_migliori_xtej1550_jets,2020_espinasse_maxji1820_jets}. X-ray light echoes generated by bright transient objects such as XRBs and GRBs, when observed behind or in the Galactic plane, offer an opportunity to study ISM \cite{2022_constantini_ism_book} (see also Section~\ref{sec:magnetars}). The AXIS large field of view and stable PSF will enable detailed characterization of the morphology and intensity evolution of these light echoes, providing unique X-ray constraints on the content of the ISM as the scattering is proportional to the physical properties of the dust (e.g., composition, grain size distribution). Light echoes also provide a unique means to measure the distance to an X-ray transient, at a typical accuracy of $\lesssim$ 10\% 
\cite{2016_heinz_v404cyg,2018_kalemci_4u1630,2023_vasilopoulos_grb221009a}.

\section{Fast Radio Bursts} 
\label{sec:FRB}
FRBs are the newest class of ms-duration, extremely bright ($10^{37}-10^{46}$\,erg\,s$^{-1}$) radio transients that are typically seen from sources outside of our Galaxy \citep{Lorimer+07}. Various sub-classes of FRBs have been identified and even localized to their host galaxies; yet their emission mechanism(s) and the progenitor engine(s) remain unknown \citep{Petroff+22}. Except for the one instance of an FRB detection from a magnetar (SGR~1935+2154) in our own Galaxy (FRB~20200428, which was seen with an X-ray counterpart; \cite{Mereghetti+20, Ridnaia+21}), searches for high-energy/multi-wavelength counterparts to all the other FRBs have only yielded non-detections \cite{Cunningham+19,Guidorzi+20,Scholz+20}. 

AXIS is poised to change this scenario. Subsecond incoherent synchrotron X-ray ``afterglow'' emissions are predicted from almost all FRB models and are corroborated by the FRB~20200428 event \citep{Lyubarsky+14, Lyutikov&Lorimer_16, Lu+20, Margalit+20}. The detection of this X-ray afterglow emission and its properties, such as the peak energy of the X-ray emission, its flux, and the X-ray-to-radio flux ratio, can inform us about the local plasma conditions such as magnetization, and composition. Characterizing these properties is paramount to confirm/refute models of the emission mechanism of extra-galactic FRBs.  Furthermore, deep X-ray observations of repeating FRB sources can also potentially identify the progenitor of these bursts (the currently prevailing theoretical models of repeating FRBs involve either a flaring magnetar or a jetted accreting compact object \citep{Sridhar+21}); the latter will also yield a persistent soft quasi-thermal X-ray counterpart due to the accretion disk and the surrounding ``hypernebula'' \citep{Sridhar+22}. X-ray studies of the host galaxies of FRBs (e.g., AGN fraction in FRB hosting galaxies) can also reveal a plethora of information about the formation channel of the compact objects emitting them, and assist in identifying newer sub-populations of FRBs \citep{Eftekhari+23}. 

The cosmological distances of FRBs have simply been too large for high-energy counterparts to be detected by the current generation of X-ray telescopes. The revolutionary capabilities of AXIS will enable observations of the repeating class of FRBs to constrain their properties and those of their host galaxies. Furthermore, ToO observations can be triggered on repeating FRBs that are quasi-periodic in nature, based on the onset of their active phase \citep{Chime/Frb_Collaboration+20, Rajwade+20}. For example, the current upper limit with Chandra on the ms-duration soft X-ray emission coincident with an FRB event is $<10^{47}$\,erg\,s$^{-1}$ \citep[for FRB~20121102;][]{Scholz+20}. Given the distance to FRB~20121102 of 972\,Gpc, any observation with AXIS longer than 1\,ks can start constraining the X-ray-to-radio flux and, therefore, the efficiency of various models of emission mechanisms.

\section{Sources of high-energy neutrinos}
\label{sec:neutrino}
The IceCube Neutrino Observatory, located at the South Pole, made a groundbreaking discovery with the detection of astrophysical neutrinos. In 2013, the facility observed high-energy neutrino flux in excess to the expected atmospheric background \cite{Aartsen+2013}. The observation marked the beginning of a new era of multi-messenger astronomy, where neutrinos serve as messengers of distant cosmic phenomena and sparked intense interest in identifying their sources. A couple of additional observations, a neutrino flare coincident with a gamma-ray flare of TXS~0506+056 \cite{IceCube2018a} and the discovery of a neutrino source compatible with NGC~1068 \cite{2022Sci...378..538I}, focused the community's attention to variable blazars and active galactic nuclei as possible candidates of high-energy neutrinos. Alternative possibilities, including accretion-powered ``hypernebulae'' in XRBs \cite{Sridhar+23b} and supernovae \cite{GCN.34837}, are also being pursued. Here, we discuss the role AXIS can play in multiple counterpart scenarios.

\subsection{Blazars}
The neutrino flare emitted from the direction of TXS~0506+056 marked a significant milestone in astrophysics, as it was the first time a high-energy neutrino was linked to a known astrophysical source. Intriguingly, the neutrino flare occurred in temporal coincidence of a gamma-ray flare \citep{IceCube2018a}. In fact, blazars are highly variable sources, whose study is of interest in its own right, for example, for the search for quasi-periodicities from an SMBHB (in fact, it has been suggested that TXS~0506+056 could be hosting an SMBHB which would merge in the LISA band within its mission lifetime \citep{deBruijn2020,BeckerTjus2022}, making it a source of \emph{three} messengers). However, the connection of such variability to high-energy neutrino emission is an open issue. Multiwavelegth variability and polarization studies play a crucial role in mapping the emitting regions along blazar jets, as well as revealing their leptonic or hadronic nature. In this context, guaranteeing the soft-X-ray coverage in long-term monitoring of the fainter blazars will be crucial to uncovering a possible link between high-energy neutrinos and relativistic jets.

\subsection{Active galaxies}
NGC~1068 (or Messier 77), is the prototypical Seyfert II galaxy, a type of active galactic nucleus (AGN) showing starburst activity, and is one of the brightest and closest to Earth. Its activity is powered by a supermassive black hole at the center, which is highly obscured along the line-of-sight by thick gas and dust. Such a dense and hot environment obscures the view of the nucleus in visible light and suppresses gamma-ray emission above hundreds of MeV. Hence, monitoring the innermost part of the source is only possible in other wavelengths, such as infrared and X-rays. However, infrared emission can be contaminated by emissions from star formation in the host galaxy. Soft X-rays can help distinguish between the two sources, as AGN emit X-rays, while star-forming regions typically do not. High energy neutrino production may result from the acceleration of ions, via magnetic reconnection and/or turbulence, up to relativistic regimes that would interact with disk photons producing neutrinos via the photo-meson production process \citep{2022arXiv220203381M}.

\subsection{Accretion-Powered Hypernebulae}
Hypernebulae are inflated by accretion-powered winds accompanying hyper-Eddington mass transfer from an evolved post-main sequence star onto a stellar-mass black hole or neutron star companion \citep{Sridhar+22, Sridhar+23b}. The conditions required to inflate these compact energetic hypernebulae are typically attained prior to common-envelope events, making these sources decades-millennia-long transients. The ions accelerated at the jet termination shock of a hypernebula can generate high-energy neutrinos via photohadronic ($p\gamma$) interactions with the softer quasi-thermal disk photons. Note that a sub-population of FRBs may be powered by such short-lived jetted hyper-accreting engines \citep{Sridhar+21}, and the radio-synchrotron-bright hypernebulae surrounding them could power the spatially-coincident `persistent radio source', and impart large rotation and dispersion measures onto the FRB \citep{Sridhar+22}. 

If the hypernebula birth rate follows that of steller-merger transients and common envelope events, their volume-integrated neutrino emission could explain $\sim 25\%$ of the high-energy diffuse neutrino flux observed by the IceCube Observatory and the Baikal-GVD Telescope. The large optical depth through the nebula to Breit-Wheeler ($\gamma\gamma$) interaction attenuates the escape of GeV-PeV gamma-rays co-produced with the neutrinos, thus rendering hypernebulae gamma-ray-faint neutrino sources, consistent with the Fermi observations of the isotropic gamma-ray background. Given the large accretion rates of hypernebulae, the disk X-ray photons do not emerge directly from the disk surface; instead, they emerge from the fast wind/jet photosphere at much larger radii. This reduces the effective temperature of the disk emission to $10\sim100$\,eV thus enabling hypernebulae to be candidates for ultraluminous supersoft X-ray sources \citep{Kahabka_06, Micic+22}. AXIS, with its large soft-X-ray effective area, would play a crucial role in discovering and characterizing the supersoft X-ray emitting hypernebulae, which might also be one of the significant contributors to the extragalactic background high-energy neutrino flux.

\section{Conclusions}
Building off the legacy of facilities such as Chandra, XMM-Newton, and Swift, AXIS will be a transformative facility for the study of the time-domain and multi-messenger universe. The combination of tremendous sensitivity, excellent field-averaged angular resolution, and rapid response capabilities makes AXIS the ideal X-ray telescope to complement the TDAMM landscape anticipated in the next decade. In this White Paper, we have described a broad range of such studies that can be conducted during the five year prime phase duration, many of which will be undertaken by the community via the robust Guest Investigator program. Perhaps most exciting, however, are the new discoveries that we cannot anticipate currently; in many areas, TDAMM studies are just in their nascent stages, and unexpected results will surely arise. With AXIS, the future of TDAMM is (X-ray) bright! 

\vspace{6pt} 

\externalbibliography{yes}
\bibliography{references}

\end{document}